# Enabling near-atomic-scale analysis of frozen water

A.A. El-Zoka1*+, S.-H. Kim1*, S. Deville2, R.C. Newman3, L.T. Stephenson1, B. Gault1,4+

1 Max-Planck-Institut für Eisenforschung, Düsseldorf, Germany.

2 Université de Lyon, Université Claude Bernard Lyon 1, CNRS, Institut Lumière Matière, 69622 Villeurbanne, France.

3 Department of Chemical Engineering & Applied Chemistry, University of Toronto, Canada

4 Department of Materials, Royal School of Mines, Imperial College London, London, UK

*Co-first authors: Ayman El-Zoka a.elzoka@mpie.de; Se-Ho Kim s.kim@mpie.de;

+ Corr. Authors: Ayman El-Zoka a.elzoka@mpie.de; Baptiste Gault b.gault@mpie.de

## Abstract

Transmission electron microscopy has undergone a revolution in recent years with the possibility to perform routine cryo-imaging of biological materials and (bio)chemical systems, as well as the possibility to image liquids via dedicated reaction cells or graphene-sandwiching. These approaches however typically require imaging a large number of specimens and reconstructing an average representation and often lack analytical capabilities. Here, using atom probe tomography we provide near atom-by-atom analyses of frozen liquids and analytical sub-nanometre three dimensional reconstructions. The analyzed ice is in contact with, and embedded within, nanoporous gold (NPG). We report the first such data on 2–3 μm thick layers of ice formed from both high purity deuterated water and a solution of 50mM NaCl in high purity deuterated water. We present a specimen preparation strategy that uses a NPG film and, additionally, we report on an analysis of the interface between nanoporous gold and frozen salt water solution with an apparent trend in the Na and Cl concentrations across the interface. We explore a range of experimental parameters to show that the atom probe analyses of bulk aqueous specimens come with their own special challenges and discuss physical processes that may produce the observed phenomena. Our study demonstrates the viability of using frozen water as a carrier for near-atomic scale analysis of objects in solution by atom probe tomography.

## Introduction

In recent decades, transmission electron microscopy (TEM) has undergone tremendous progress, in part leading to the 2017 Nobel Prize in chemistry to Jacques Dubochet, Joachim Frank and Richard Henderson "for developing cryo-electron microscopy for the high-resolution structure determination of biomolecules in solution"[1]. In parallel to these cryo-TEM developments in the biological sciences, there has been developments around liquid cells for imaging nanoscale objects [2,3]. The early studies all highlighted where the core difficulties were: on how to make and transfer specimens that maintained the sample in its pristine state in a completely different environment, i.e. in the vacuum of the TEM, while enabling imaging of embedded objects. Tremendous efforts, in parallel, have established atomically resolved electron tomography that led to ground-breaking results on crystal defects[4], order/disorder transitions[5] and nucleation events[6]. Albeit powerful, these approaches have limited analytical capabilities and hence cannot easily measure the specimen's atomic-scale composition.

Atom probe tomography (APT) is a burgeoning technique that enables 3D elemental mapping[7] with sub-nanometre resolution[8] and a chemical sensitivity potentially down to only tens of parts-per-million. The usefulness of atom probe tomography (APT) to study wet chemical systems has been limited by the challenges inherent to the analysis of liquid media. There are ongoing efforts to push the development of "cryo-APT" – even though the term is slightly inappropriate as all APT experiments are performed at cryogenic temperature, typically at 20-80K. The development of cryo-approaches for specimen preparation and specimen transfer[9–11] have enabled some steps towards the analysis of limited volumes of frozen water[12–15] which can be sandwiched in between a needle-shaped metallic specimen and graphene sheets[16], akin to the liquid cells for TEM[3]. To pave the way towards enabling these studies on large volumes, strategies for specimen preparation and systematic studies are necessary to enable precise analysis and assess the performance limits of cryo-APT. This could unlock a more widespread application of APT to study individual molecules – once referred to as a 'chimera' by a pioneer of the technique[17]. There are sporadic reports of using APT for investigating molecular and biological materials[18–25] and the use of APT nano-scale objects, i.e. nanoparticles[26–29], nanowires[30–33] or nanosheets[34], but these samples are always taken out of their native or *in operando* environments and dedicated strategies had to be developed to enable specimen preparation[35,36,37,38].

The analysis of 'bulk' ice, similar to that performed on a metal, is yet to be carried out. The importance of ice and water, in its numerous physical states, is not limited to the fields of physics and chemistry. It is also relevant to fields such as biology [22], atmosphere chemistry[39], geophysics[40], and even space studies[41]. The problem of the low electrical conductivity of bulk ice caused the failure of the analysis of a layer of ice tens of microns in thickness[42]. Recent notable attempts showed the possibility of analysing hydrated glass by APT [11].

Here, we report on the analysis of micron-thick layers of frozen water formed on nanoporous gold (NPG). NPG is formed by the selective dissolution of Ag from a AgAu solid solution in acidic conditions, causing the formation of a 3D, bicontinuous open-pore nanostructure[38,43]. NPG has found application in catalysis, electrochemical sensing and actuation, owing to a high surface area-to-volume ratio and Au-rich surfaces. In this work, we take advantage of NPG as a hydrophilic substrate to analyze ice by APT. NPG typically exhibits hydrophilic behaviour, and wetting increases with decreasing pore size[44,45].

To prepare specimens suitable for field evaporation within the state-of-the-art atom probe microscope, we adapted the blotting and plunge-freezing approaches typical from cryo-TEM. We employed an in-situ plasma-focused ion beam[46] approach, albeit here at cryo-temperature. The low contact angle of the water droplet on the NPG substrate facilitated the preparation of stable specimens fully comprised of frozen liquid. The complex and continuous network of pores also provided strength to the ice-substrate interface. We report on a wide range of pulsed-laser atom probe data from pure deuterated water ($D_2O$) and a solution of 50 mM NaCl in $D_2O$, obtained by systematically sweeping through some of the main experimental parameters. We also demonstrate the ability to characterize small metallic objects floating in solution and the ice/solid interface, by analyzing data captured at the ice/NPG interfaces. We discuss aspects of the physics of field evaporation that lead to the detection of sets of molecular ions and their influence on the performance of cryo-APT. Our study is a necessary step towards opening a new playing field for near-atomic scale analysis of solute effects in confined freezing, nano-objects, and molecular or biological materials in their native environment.

## Results

*Specimen preparation*

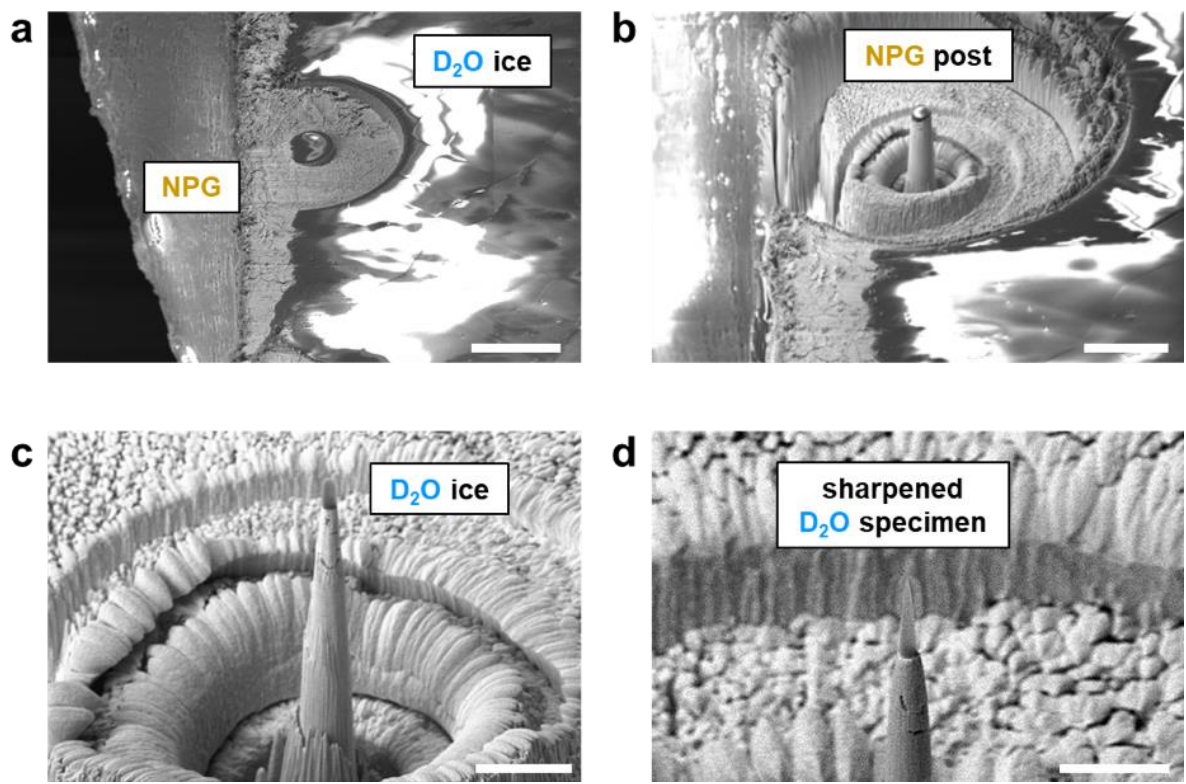

**Figure 1- SEM images of in-situ APT specimen preparation of ice sample on NPG: (a) 200 and 75-μm ion beam annular pattern for outer and inner diameters, respectively, were made on the ice/NPG sample, scale bar is 100 μm. (b) The ice/NPG pillar was milled until the height of the Au post reached >50 μm[47]. Scale bar is 50 μm. (c) Ice layer was gradually sharpened along with NPG Au until the layer reached <5 μm in height, scale bar is 15 μm. (d) Final APT specimen of ice on NPG, scale bar is 5 μm.**

APT requires needle-shaped specimens to produce an electrostatic field with sufficient intensity to provoke field evaporation of the surface atoms. With the availability of plasma focused ion beams (PFIB), an *in-situ* APT specimen preparation strategy was revisited recently[46,48], in which specimens were fabricated directly from the original flat sample prepared for scanning-electron microscopy (SEM) imaging itself, eliminating the need for conventional liftout and Pt-welding procedures[49]. The details of each step are described in the Methods section and outlined in Figure 1. The low contact angle of the water droplet on the NPG substrate facilitated the preparation of stable specimens comprising the frozen liquid and the metallic interface. The cross-sectional image of the interface structure of ice-on-NPG substrate shows that there are no noticeable voids along the interface or within NPG substrate (see the Supplementary Information Figure S1). Several specimens were made to analyze the ice at different distances from the NPG layer. The difficulties associated with ion milling of large volumes of ice were mitigated in these experiments by aiming at the substrate edges, where the thickness of the ice layer is at its minimum, due to the wettability of the NPG.

We cannot measure the cooling rate using our current protocol and so whether ice is crystalline or amorphous is uncertain. In the thicker regions of the frozen droplet, the crystalline nature of the ice was confirmed by SEM during the preparation of the specimen as grain boundaries were observed in the ice on top of the NPG (Figure S2). A rough estimate of the cooling process (see methods) indicates that the sample might have taken ~0.2s to cool from room temperature to 136K, i.e. one-to-two orders of magnitude below the approximately $10^4$ $Ks^{-1}$ necessary for vitrification at ambient pressure. Amorphous ice would be necessary for the cryo-preservation of solutions or of biological materials, yet the quenching process implemented here with liquid nitrogen likely only yields crystalline ice and experiments using liquid ethane-propane mixture or other cryogens will be necessary in the future. Preparation in the PFIB can lead to amorphization associated with damage from the incoming energetic ions [50], even though FIB-damage to ice thin sections was previously shown to be limited [51]. An experimental proof of the crystallinity of the APT specimens could not be obtained – this likely requires the implementation of a correlative protocol involving cryo-transmission-Kikuchi diffraction or cryo-TEM and cryo-APT, which is currently out of reach. Heavy water-based solutions are used, to distinguish between hydrogen in the frozen liquid from that found as residual gas in the analysis chamber [52].

*APT of ice*

We managed to overcome the challenge of analysing a relatively thick (~ 2 µm) layer of ice, which might appear surprising considering the intrinsic highly insulating nature of crystalline ice. Previously, the analysis of water layers by field evaporation was limited to nanoscale thicknesses of water layers condensed on top of metallic needles suitable for field evaporation [53] and where the electrostatic field could hence be generated. It is then no surprise that the discussion so far has largely focused on the first contact layer between water and metals. However, there is very little known about the structure of water beyond the few first few layers [54]. The electrostatic field at the surface of insulating materials analysed by APT, for instance MgO, was thought to be generated thanks to the bending of the electronic bands combined with surface defect states enabling absorption of the laser light[55,56]. Such aspects have not been reported for thick ice layers.

Using a combination of in-situ APT preparation and protocols for the transfer of environmentally-sensitive specimens, we repeatedly collected data displaying ice chemistry at near-atomic resolution. A summary of a dataset with >40 million molecular ions of $D_2O$ with 3D atom map is presented in Figure S5. These data were acquired without any observed microfractures of the specimen tip, as often happens with apparently fragile specimens, and instead showed a smooth evolution of the applied DC voltage over the course of the experiment.

Figure 2a shows a typical mass spectrum obtained from the pulsed-laser APT analysis. The general trend we observed when it comes to the detected cations was that the ice evaporates in the form of singly-charged molecular ions of 1–5 $D_2O$ molecules. We also detected such water clusters protonated with H and D atoms interchangeably; however fully deuterated clusters always dominated other cluster groups in abundance. The proton-donating ability of water can lead to many other protonated inorganic ions during field desorption as previously reported[57]. Cluster formation is believed to be due to ion aggregation and polymerization prior to field evaporation. The formation and subsequent possible dissociation of chains of water molecules and molecular ions under the influence of an intense electric field were predicted by density-functional theory[58], and our results qualitatively agree with these predictions. These aspects will be further discussed below. We used the default reconstruction protocol in the commercial software package IVAS to build a point cloud with individual points being an individual molecular ion – the volume of each ion being the default volume of the *n* oxygen atoms within the molecular ion ($0.0288 nm^3$). In this case, default reconstruction parameters were used making the overall data scaling arbitrary. Yet, one can assume that the relative topology of the data on a local scale would map relatively faithfully to the structure in the original specimen, even though the effective spatial resolution in such a case is not precisely known[8]. The tomogram and scanning electron micrograph of the corresponding specimen (inset) are shown in **Error! Reference source not found.**c.

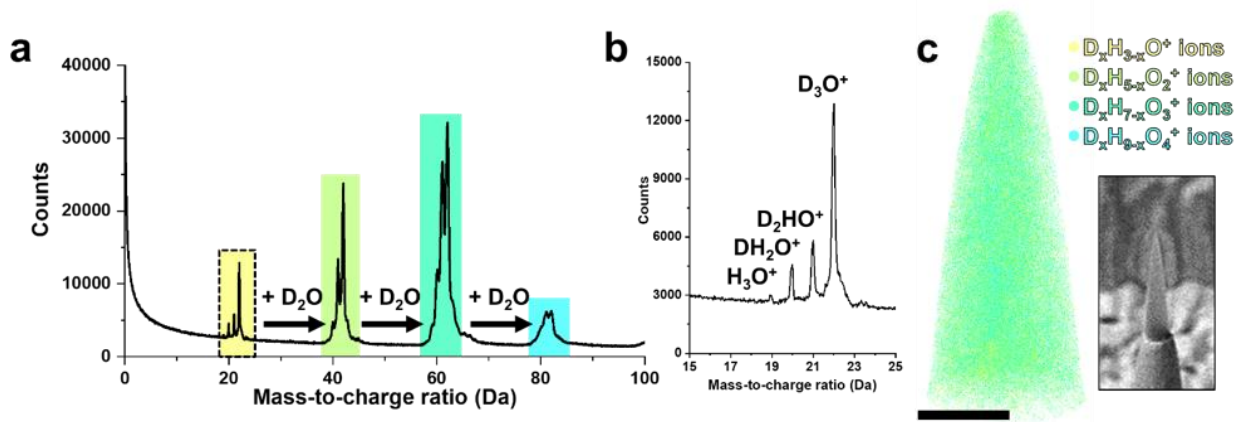

**Figure 2-** (a) Mass spectrum of acquired APT dataset of $D_2O$ ice at 100 pJ, 200 kHz and a detection rate of 0.5 %. (b) Sectioned mass spectrum from Figure 2a to illustrate $D_xH_{3-x}O$ complex peaks. (c) 3D reconstruction map of $D_2O$. Inset capture shows e-beam image of the specimen. Scale bar is 50 nm.

*APT of embedded nanoscale objects and NPG/ice interface*

Our preliminary work also shows the possibility of analysing frozen liquid-metal interfaces. The dealloying process produces Ag ions within the vicinity of the nanoporous layer. Areas within the ice layer were found to contain agglomerations of Ag ions, likely solvated silver floating away from the NPG substrate. Figure 3 shows the accompanying mass-to-charge spectrum, where hydrated Ag ions were detected at low pulsing energies (20 pJ). This reveals that there are indeed remnant Ag ions close to the nanoporous layer, even after a day of immersing the sample in water. The composition profile calculated through one of the Ag flakes shows that the Ag composition locally reaches nearly 90 at. %, yet still contains $D_2O$ ions at the core which could be due to trajectory aberrations. However, the detection of molecular ions mixing Ag and water points to the fact that these Ag structures are in a solvated aqueous state. Our approach hence enables analysis of nanostructures embedded in ice.

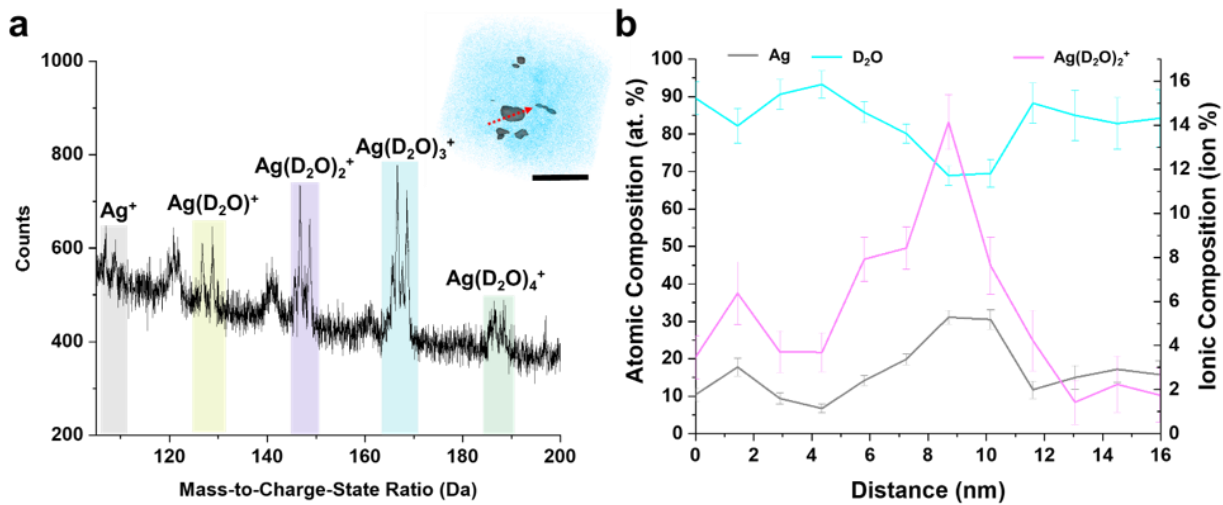

**Figure 3-** (a) Ag+ ions observed within the ice layer, inset showing group of Ag+ ions surrounded by ice with an isosurface value of 10 at. % for Ag. (b) Chemical profile across Ag+/D$_2$O interface shown by red arrow in (a). Scale bar measures 20 nm.

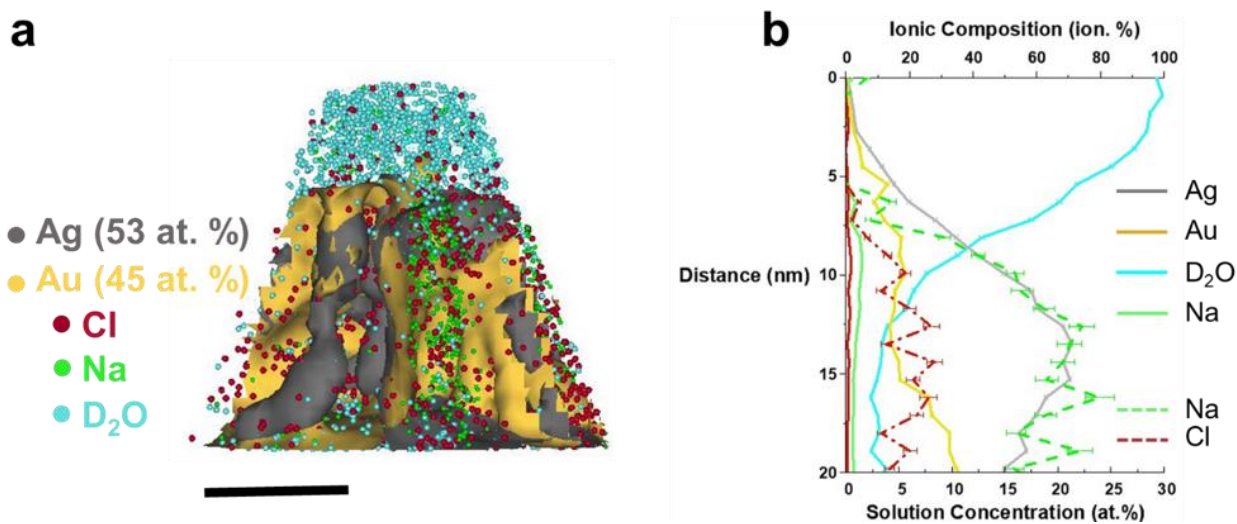

**Figure 4-** (a) 3D atom map of the ice-NPG substrate interface and a corresponding chemical map, highlighting the concentrated presence of Na inside the nanopores rather than away from the NPG. Scale bar is 10 nm. (b) Atomic concentrations along the z-axis from the 3D atom map in Figure 4a.

We can also report on the ice embedded within the highly porous network of Au-rich nanoligaments. Figure 4 demonstrates the capability of analysing the ice/NPG continuously. Important questions arise regarding the effects of size confinement on ice when considering the water inside the nanopores, as impact on phase growth[59], solute distribution[60,61], and electrical conductivity are expected[62]. We can here map the chemical composition across the water/ substrate interface, revealing the high concentration of Na+ and Cl- ions in the nanoporous layer. The cations and anions have a high affinity to the NPG surfaces. The composition profile in Figure 4b, calculated across the entire cross-section of the analyzed region, shows an increase to up to approx. 5 at. % Na near NPG surfaces. Na+ and Cl- were only detected in close proximity to the NPG layer. Experiments performed in the hundreds of nm to several μm range above this interface showed no signs of Na or Cl. This can be explained by the cooling rate being too slow to freeze the cations/anions in place during the quench. Thermodynamically, the nanopores are expected to be the last sites to freeze, so the growth of the ice from the surface of the water layer towards the pores layer pushes the Na+ and Cl-ions into the pores. This was seen previously in studies focused on the solute concentration distribution in NaCl solutions and others, but never on such a scale.[63,64]

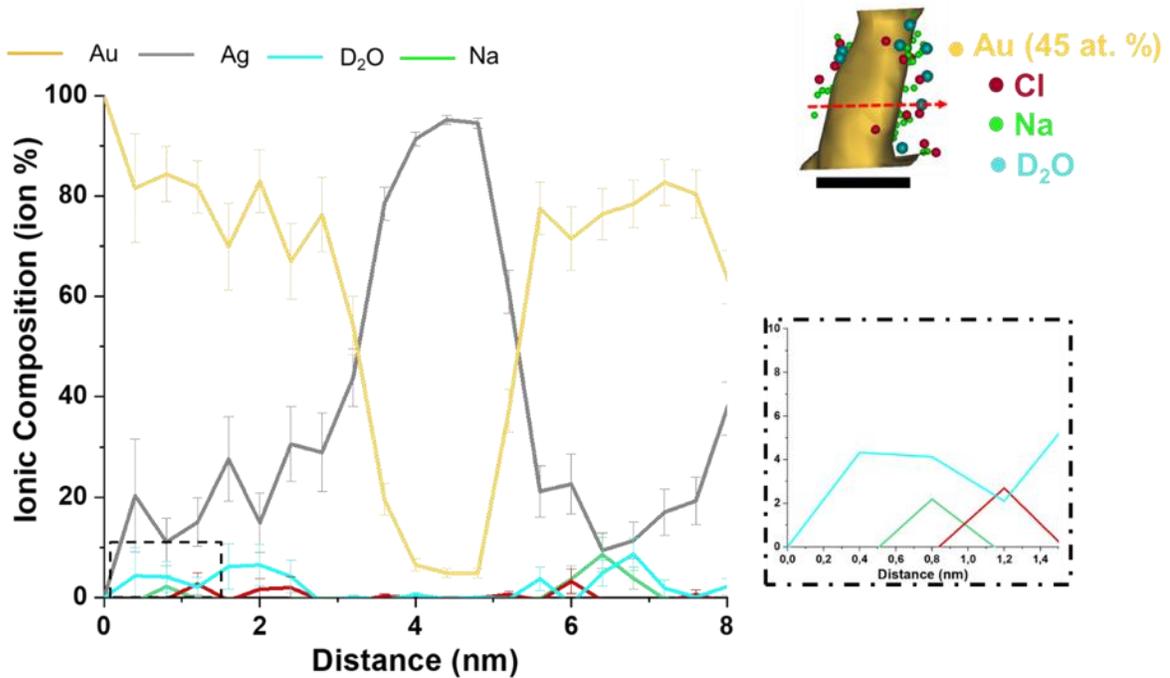

**Figure 5- Ice-nanoligament interface showing the high concentration of Na ions near the Au surfaces, and as seen before[38], the core-shell structure of the nanoligament. Inset showing a the mapped nanoligament and the surrounding ions detected. Scale bar is 5 nm.**

In addition, we were able to isolate one of the several imaged nanoligaments within the nanoporous layer. As we expect from previous APT studies[38,65] on NPG, nanoligaments show a Ag-rich core, and Au-rich surfaces, attributed to the surface diffusion and agglomeration of Au atoms during the selective dissolution of Ag atoms. The atom map in also shows an enrichment of Na and D$_2$O ions close to the ligament/ice interface. Na$_+$ ion was detected around the nanoligaments as the hydrated Na(D$_2$O)$_+$ ion, in addition to the elemental form. Hydrated metal ions are not an anomaly, as they were observed with Si in previous field evaporation studies on adsorbed water layers[66].

For the case of Au and Ag, hydrated and hydrogenated clusters are observed, possibly coming from the ice/NPG interface[57]. The composition of Ag at the ligament core is higher than might be expected (77 at. %), as the surface diffusion-based model of the dealloying process supports a composition at the ligament core close to the bulk composition of the parent alloy. At a heterogeneous interface such as the one shown in

, trajectory aberrations are expected to play a role, especially when it comes to the compositional estimates – this may explain the relatively low composition of water around the ligaments as atoms originating from the ligaments are subject to strong aberrations and are imaged as part of the pores, typical of the aberrations caused by differences in the required electric field to provoke the field evaporation[67,68]. Further experimentation will help optimize conditions to maximise compositional accuracy. Nevertheless, the results presented here showed the successful analysis of a nanostructure embedded in ice, and this could be further translated into embedding other composite nanostructures of interest.

*Influence of acquisition parameters on APT data*

Here we highlight the outcomes of our first foray into how the main experimental parameters influenced the collected APT data for the frozen aqueous samples. APT experiments are typically performed at a constant detection rate, which is maintained by the acquisition software's control algorithm. The standing electrostatic field (controlled by the applied standing voltage) is the main factor effecting the detection rate, but the detection rate is affected by other user-defined parameters. For example, the pulsed field strength in HV pulsing (controlled by the pulse fraction, a percentage of the standing voltage) or the thermal pulse response (controlled by the laser pulse energy). As either of these user-defined parameters are decreased, the evaporation rate and thus the detection rate decreases. In response, the control algorithm compensates by increasing the standing voltage, so increasing the standing electrostatic field. This compensates for the lower field or temperature reached by the specimen's surface following the HV pulse or laser light absorption by the specimen. For laser pulsing, this leads to a significant change in the relative ratio of molecular ions detected. Lower pulsing energies lead to a higher electrostatic field, and this favour small clusters, while higher pulsing energies lead to a lower electrostatic field, and this resulted in an increase of larger molecular ions (m = 4 and 5). Figure 6 shows the calculated cluster concentrations at the different pulsing energies attempted. Previous work examining field evaporation ionization of water showed that lower fields favour larger clusters[69]. This trend could be changed by the base temperature of the specimen during analysis and a temperature sweep experiment reported by Stintz and Panitz[70] was shown to affect the order of cluster abundance. Furthermore, a high-voltage pulsing analysis of an ice specimen showed a considerable shift in ion abundances towards the dimer cluster, confirming the trend.

Separate oxygen and hydrogen/deuterium peaks from the dissociation of water were detected at low laser pulsing energies and in HV pulsing. Dissociation of water was also observed in close proximity to Ag/Au ions, i.e. close to the ice-metal interface where the strength of the electrostatic field was also higher. These conditions are closer to those in previous reports of the analysis of ice by high-field techniques. The detection of elemental peaks in the case of ice depends on the intensity of the electrostatic field, consistent with previous observations by Tsong and Liou [71], who showed through field desorption experiments that protonated water clusters only dominate at lower electrostatic fields (higher pulsing energies). This is similar to most reports of molecular ions and their relative stability with respect to the intensity of the electrostatic field [72–76].

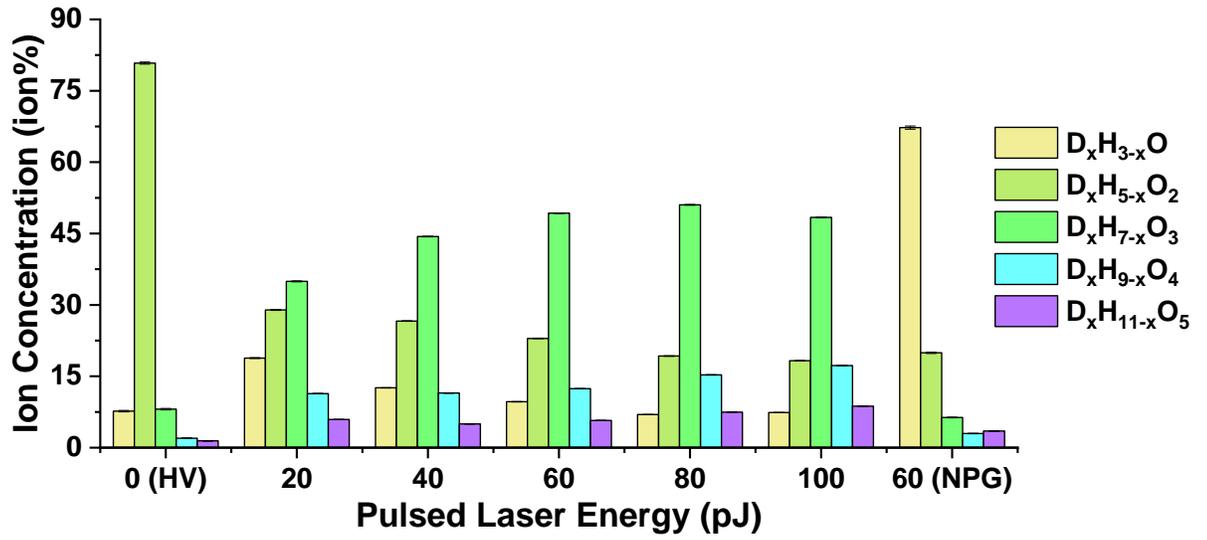

**Figure 6-** Ion concentration of different clusters at pulsing energies ranging from 20-100 pJ. Pulsing rate for the HV measurement was 15 %.

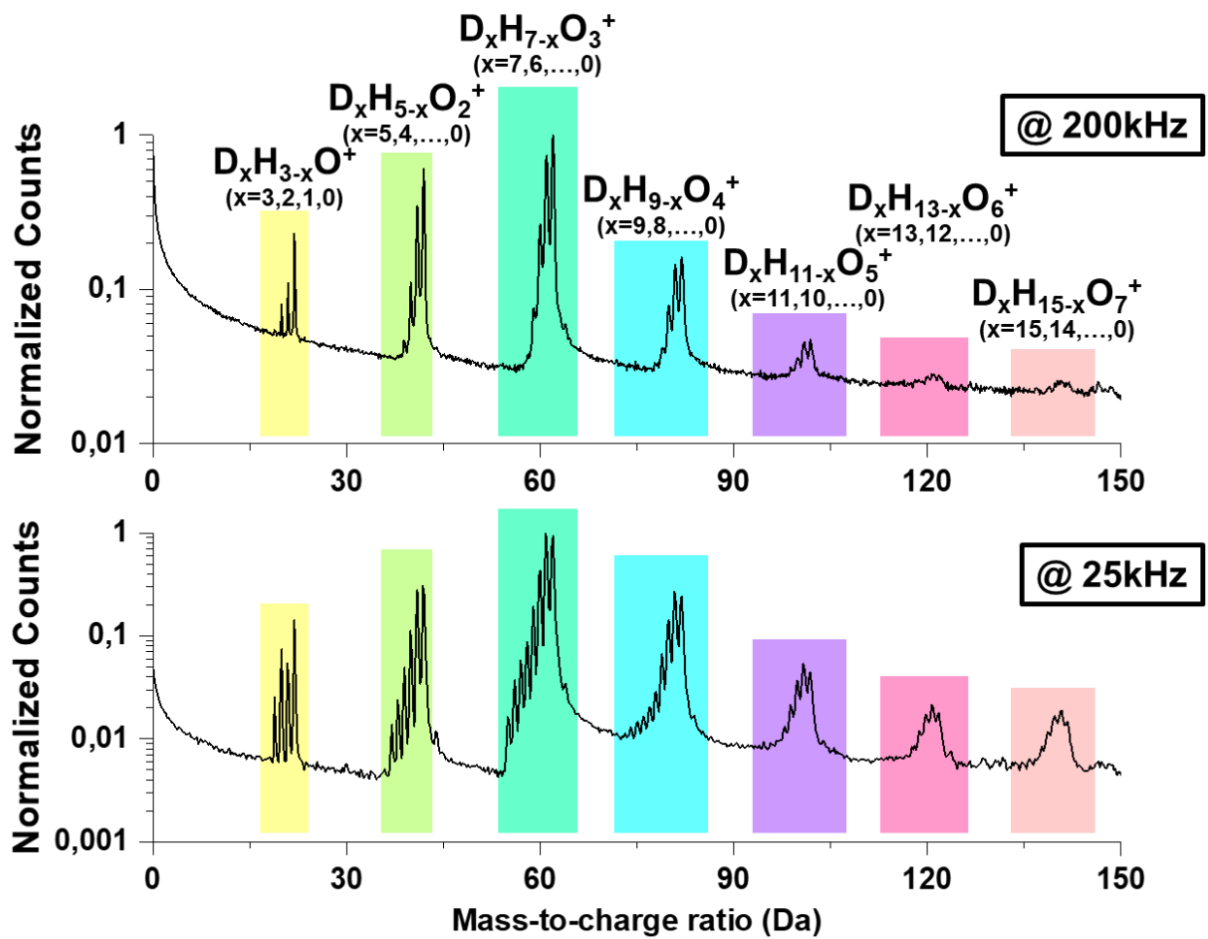

**Figure 7** - Mass spectra from the NaCl-$D_2O$ water specimen at 200 and 25kHz – note the difference in scale on the y-axis.

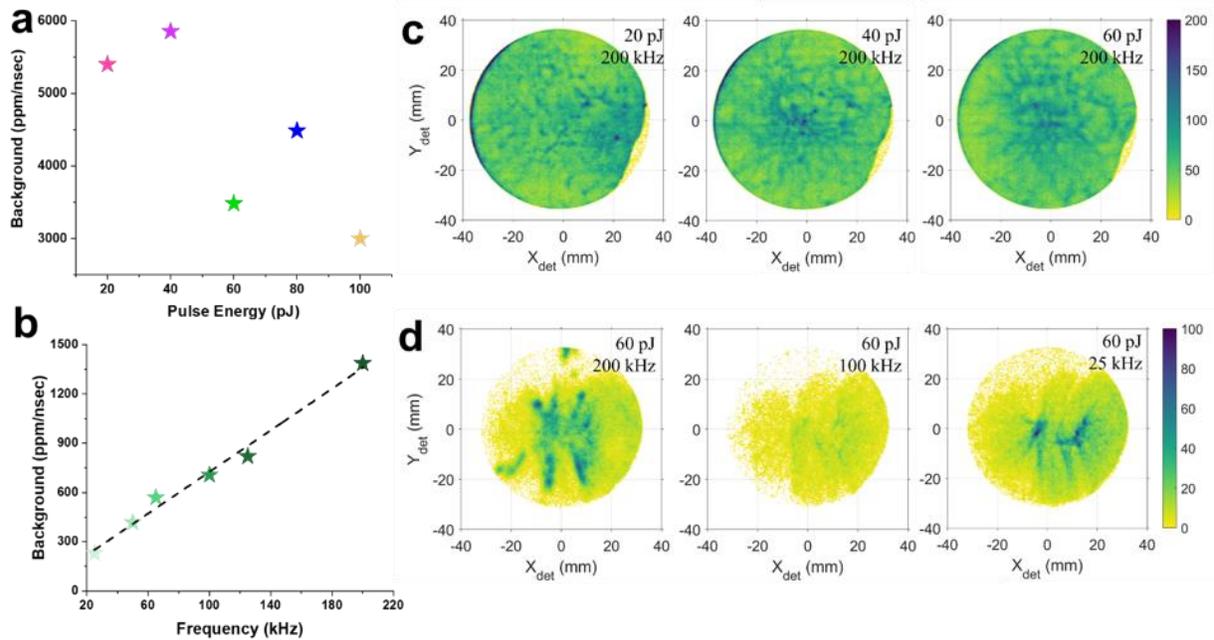

**Figure 8** – (a) level of background as a function of the laser pulse energy at 200kHz pulsing rate and effect of the pulsing rate at 60pJ. (b) two-dimensional detector hit maps for a range of parameters.

It is critical to quantify the level of background in order to assess the sensitivity of the APT analysis of solutions. The background levels detected in most experiments were relatively high in comparison to the usual analysis conditions for APT but could be lowered by changing the experimental parameters. Ice being a notoriously poor heat conductor, we lowered the repetition rate of the laser in order to potentially avoid a pile-up of the thermal pulses, which indeed led to a drop in the background in the analysis of salt water and an enhanced signal-to-background ratio, as shown in Figure 7. In Figure 8a, we report on the estimated background level in a series of analyses as a function of the laser pulse energy and at a 200kHz pulse repetition rate. The estimation is performed according to the procedure implemented in the commercial software Cameca IVAS 3.8.4. The background level was not significantly reduced by lowering the pulsing energy at 200 kHz. Figure 8b shows that for the salt water specimen, we systematically changed the laser pulsing repetition rate finding a positive linear correlation. The laser pulsing rate controls how much energy is supplied to the ice tip per second, corresponding to an increased average temperature. The less energy per second provided, the more controllable is the ice field evaporation, and the lower the detected background. Figure 8c and d feature the two-dimensional detector histograms for different experimental parameters – varying pulse energy and pulse frequency respectively – showing that the homogeneity of the field evaporation process increases with decreasing pulse energies. In the supplementary, Figure S9 shows that the intensity distribution of the detector histograms for a higher pulse energy are slightly narrower. This more uniform distribution could be associated to a smoother specimen curvature. The pulse frequency and detection rates seemed to have a limited effect on the map's homogeneity.

*Aspects of the APT analysis of ice and performance limits*

Our preliminary results show that using water as a carrier medium to enable the analysis of materials will require fine tuning of the experimental conditions so as to maximise the signal-to-background-ratio in particular around mass-to-charge ratios where ions of interest are expected to be found. It is critical to minimise the background level in order to maximise the sensitivity of the APT analysis of aqueous solutions. The issue of detection sensitivity in APT is still not well defined and there is not a specific quantitative metric to describe how low a composition can be measured. A metric for the compositional sensitivity should combine the mass resolution, i.e. how high a mass peak is for a given number of detected ions, the number of stable isotopes that lead to lower peak amplitude potentially, but also the local level of background that could obscure this specific set of peaks. We provide here an estimate for the case of Na ions in a deuterated solution. The background level at 23 Da, where the peak for the single stable isotope of Na is located, is 490 counts. If we assume that to enable the detection of Na, we need to detect at least a number of ions equivalent to the level of background, i.e. approx. 500 counts within 1 million ions in this particular range, then this translates to a sensitivity of approx. 0.05 at% which can be translated into approx. 28mM. This value of the sensitivity would matter in the case where the plunge freezing had allowed for vitrification and Na was homogeneously distributed within the frozen solution. Here, the Na was highly concentrated near the NPG-liquid interface and within the nanopores, facilitating its detection. The present set of data will help guide future optimisation of the experimental parameters to maximise sensitivity for specific elements.

*Field-evaporation mechanism*

The success of our experiments might be initially considered as counterintuitive, as Stuve predicted[14] that the applied field needed to initiate ionization increases as a function of the increase in ice layer thickness, as a result of the dielectric screening by the water layer, thicker water layers being more effective at screening the field at the water−vacuum interface. Experiments and modelling in their work show the electrostatic field to be concentrated at the interface between the needle-shaped specimen and the water. The difference in our experiment was that our entire specimen was made of ice leading to the conclusion that an electrostatic field must be concentrated at the ice-vacuum interface throughout the experiment. This is summarized in Figure 9. The expected electrostatic field intensity is also much lower in our case than in other recent reports where the field is sufficiently high to lead, typically, to the fragmentation of the water molecules[16,77].

It was pointed out by Perea et al.[9], that the mode of specimen transfer and experimental conditions influence the nature of the molecular ions detected, and that it was potentially possible to detect even long chains singly-ionized chains, suggesting their controlled field-induced desorption and ionization. This is an important consideration, since there is little known and understood about the dissociation process of long, complex carbon chains found, for instance in the field evaporation and APT analysis of polymers and biomolecules[21,78–80], making their reconstruction and reconstitution, if not impossible, at least extremely challenging. Conversely to other recent reports where the analysis of liquids was performed on thin layers of ice formed on very sharp needle-shaped specimens and hence subjected to very intense fields, our approach enables a gentler field evaporation, with the hope that complete molecules or only mildly fragmented molecules could be detected.

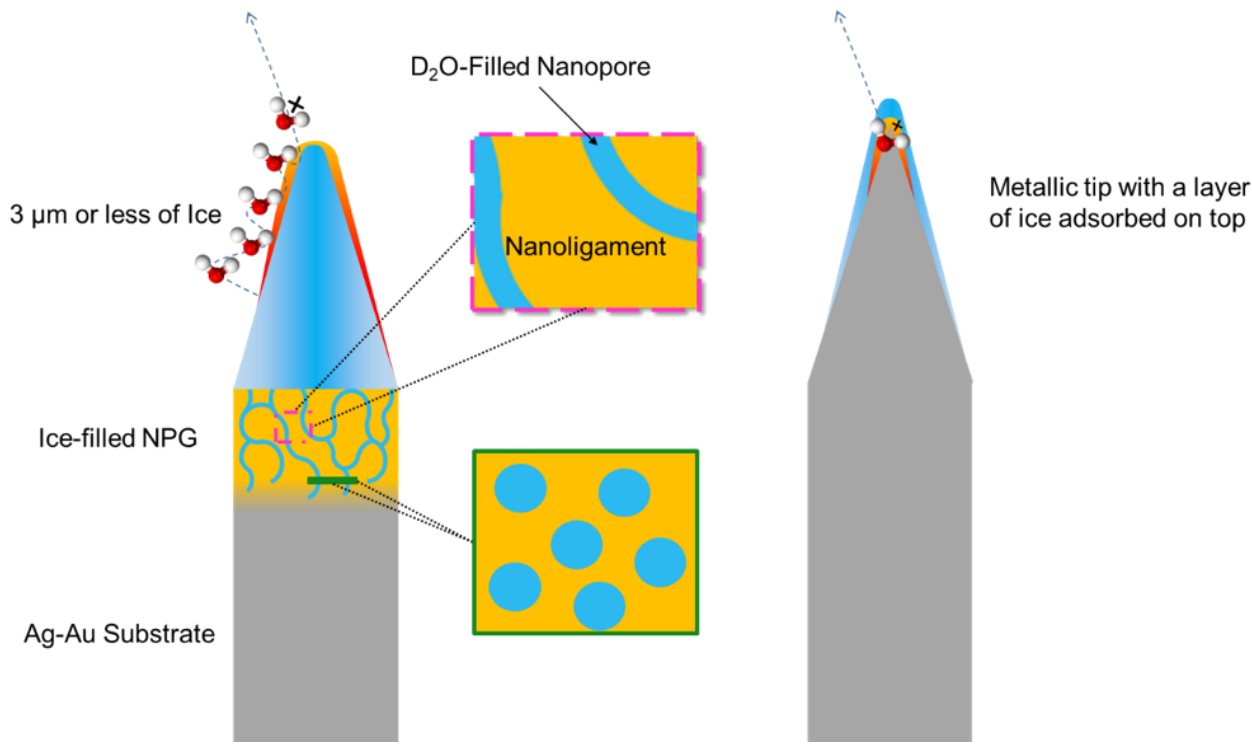

**Figure 9 - Schematic showing a comparison between the experimental setup used in previous field evaporation on ice studies, and the setup used in this work.**

The current data might not allow us to pinpoint precisely which is the most prominent field evaporation mechanism active during laser-pulsed or high-voltage induced field evaporation. There are several possible mechanisms that can lead to the detection of singly-charged monomer or polymer of water molecular ions. First, as in the case of metals, a partially charged molecule can be progressively repelled from the surface under the influence of the electrostatic field, undergoing simultaneous ionization and desorption[81]. Second, there is a possibility that thermally-activated localized migration processes lead to a rearrangement of the specimen's surface whereby several water molecules combine to form a polymer. This process is assisted by the electrostatic field, and was predicted by atomistic calcualtions[58]. These very local processes could explain the detection of the larger molecular ions detected in our experiments.

Finally, we postulate the presence of liquid-like outermost layer that could very well be partially charged, acting as a medium between the bulk ice and the vacuum. The increase in temperature subsequent to the laser pulses provides thermal agitation that would facilitate the formation of a layer comprising loosely bonded, mobile molecules. A similar layer was reported in liquid metal ion sources, albeit at higher temperatures than what we observe here [82]. Individual molecules or polymers migrate under the influence of the electric field itself, via polarization forces, to the loci at the specimen's surface where the electrostatic field is sufficiently intense to provoke their field evaporation. A similar behaviour of solutes in metallic alloys have sometimes been reported as well[83]. This hypothesis, shown schematically in Figure 10, explains the pattern formed on the detector (Figure 8b) as well as the high level of background, at least in part.

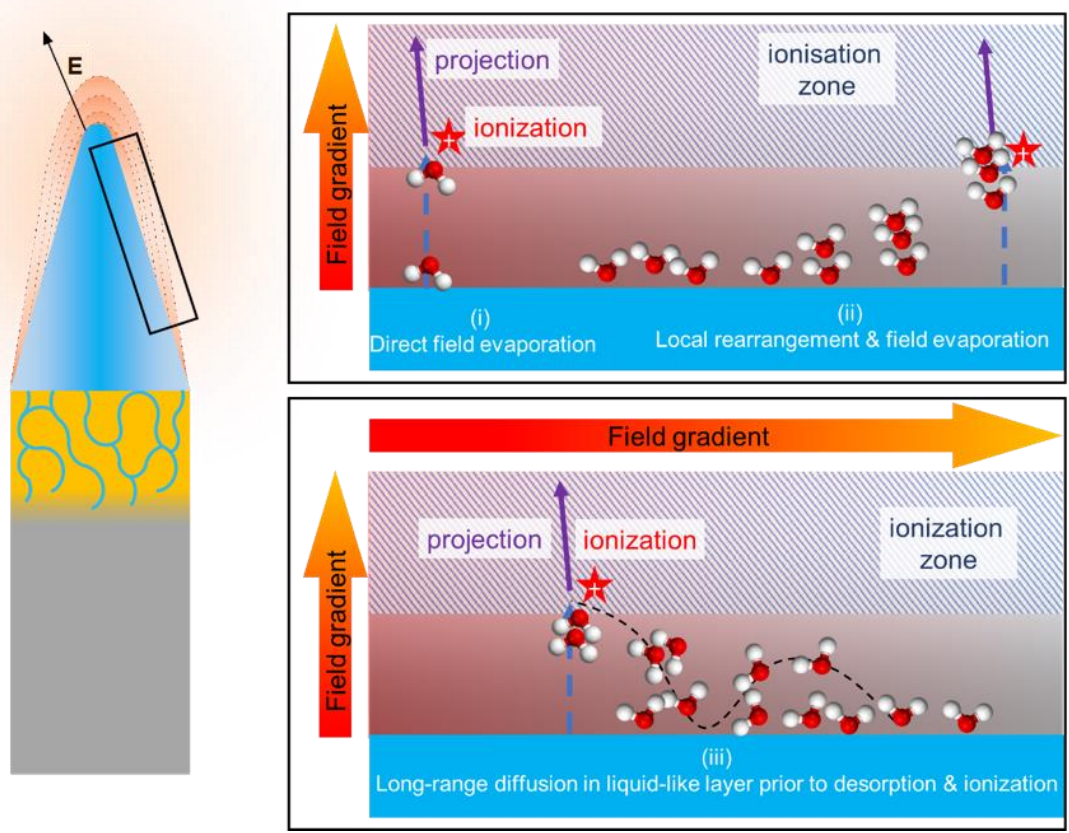

**Figure 10- Schematic showing the main steps involved in the proposed mechanism for laser-pulsed field evaporation of ice.**

The background can be related to the loosely bonded molecules that leave the specimen at a time uncorrelated with the laser pulse. The surface gets stabilized by the field desorption and ionization or field evaporation of the least stable surface atoms or molecules[84,85]. This implies that following the removal of a molecule, the rearrangement of the surface can lead to molecular migrations, and potentially, at a later stage, the removal of more or more molecular ions. This is evident in Fig S11(a) where vertical and horizontal trails co-evaporate with the pulse-correlated water cluster peaks, but there is also continuous evaporation evidenced by faint diagonal lines. This additional contribution to the background could be from water vapour molecules that undergo ionization as they desorb from the specimen's surface [86]. The low vacuum in the analysis chamber makes the evaporation of ice, in theory, easier at lower temperatures[87]. The low barrier for ionization of ice is calculated to be 0.75 eV, compared to 0.5 eV for neutral water, which might justify the ease of the ionization of water[69]. Another striking but faint feature of Figure S10 (a) is dissociation of water clusters leading to the formation of neutral water and perhaps suggesting that under some circumstance significant amounts of water may go undetected. Figure S10 (b) however suggested that the evaporation of confined water can be quite different and there we observed that numerous metallic species, along with the expected salt ions, can be filtered and characterised by examining multiple evaporation. This sensitivity of this technique can be then pushed beyond what could be expected from merely examining the straight-forward mass-to-charge spectra and the accompanying high background.

## Conclusions

The approach introduced herein has overcome the barriers faced by conventional FIB/APT analysis of liquid layers and nanostructures encapsulated in liquid layers. We demonstrated the use of NPG as a substrate for making ice needles, using a cryo-PFIB that are suitable for atom probe analysis. The results in this investigation demonstrate for the first time, the capability of analyzing bulk ice layers, and their use in probing encapsulated nanoligaments and solvated ions at the near atomic-scale. Our approach paves the way for using nanoporous metals for routinely investigating liquid layers and encapsulated nanostructures. An important optimization to make in this regard is the chemistry of the nanoporous metal and the pore size. Thermal coarsening of nanoporous metals to control feature size is possible, as was shown by El-Zoka et al.[88] through thermal coarsening studies. Replacing Au with other metals, with lower evaporation field, could eliminate some of the aberrations at the ice/solid interface and within the nanopores. Attempting other mesoporous and nanoporous systems with different feature sizes can lead to numerous studies in their own right – in particular on the influence of pore size on the local solubility, but also the composition of the layer near the surface of (for example) catalysts. Within the ice layer itself, there will be opportunities for analysing the distribution of solutes, and the early stage of formation of precipitates, including the possible role or incorporation of impurities [32,34]. Our set of experiments is a first yet major step forward towards enabling near-atomic-scale analytical imaging of chemical, biochemical and biological systems.

## Acknowledgements

Prof. Julie M. Cairney from the University of Sydney is acknowledged for fruitful discussions that led to us thinking of using NPG as a support. We thank Uwe Tezins, Christian Broß and Andreas Sturm for their support to the FIB and APT facilities at MPIE. We are grateful for the financial support from the BMBF via the project UGSLIT and the Max-Planck Gesellschaft via the Laplace project. A.E., S.-H.K., L.T.S. and B.G. acknowledge financial support from the ERC-CoG-SHINE-771602.

## Methods

**Dealloying.** A 1-cm$^2$ sample of AgAu foil was polished and annealed for 1 hour at 900 $^o$C in an inert Ar atmosphere. Then the sample was immersed in 10 ml solution of 65 % nitric acid for 5 mins. The dealloying was then stopped by transferring the sample to a solution of $D_2O$ (Sigma-Aldrich (Germany), 99.9 at.% for D). Sample was then quickly mounted on a Cu APT clip, making sure the sample is never dry, and left immersed in $D_2O$ overnight (see Figure S6). For another set of samples, the procedure above was repeated for NPG immersed in 50 mM NaCl-$D_2O$ solutions (Sigma-Aldrich (Germany), ACS reagent, 99.0%).

**$D_2O$ Treatment and Freezing.** Freezing of the sample was carried out inside a nitrogen glovebox, supplied with a constant flow of dry liquid $N_2$, to limit the formation of frost on the sample. Dew point inside the glovebox was kept as low as -99 $^o$C, and the oxygen levels were kept below 1 ppm. Freezing of the sample was done by swiftly removing the immersed clip from the $D_2O$ solution, followed by slight blotting of the surface using a kimwipe, to avoid large volumes of ice accumulating on top of the surface, and then directly immersing the whole clip onto dry liquid $N_2$ for 5 mins. Then, the sample is quickly loaded in a cryogenic ultra-high vacuum (UHV) suitcase and transferred to Xe-plasma focused ion beam (PFIB) for APT specimen fabrication.

**Cryo-FIB equipment.** All details on this specific setup forming the Laplace Project at the Max-Planck-Institute für Eisenforschung GmbH are reported in ref.[10]. It revolves around a dual-beam scanning electron microscope/focused ion beam (SEM/FIB) FEI Helios PFIB with a Xe-plasma source. A custom solid-state cooling stage connected to a dewar and a cold finger are fitted. The cryo-stage is isolated from the SEM by a series of vacuum polyether ether ketone (PEEK) spacers and accommodates a commercial puck that carries APT specimens.

The protocol introduced by Halpin et al. [46] was adapted to dig a moat and leave a pillar in the middle. The pillar is then sharpened with a Xe-plasma beam. Milling currents ranged from 1.3 µA down to 0.1 µA at 30kV. The ice specimen length was ensured to be no more than 5 microns (see Supplementary Information).

**Cooling Rates.** We cannot currently state any specific experimental values as to how fast the cooling is, this is subject to future theoretical and experimental work. We could however, propose a higher and a lower estimate for our cooling rates through simple calculations that assume the heat transfer to be a 1D problem, and that the freezing begins at the water/liquid nitrogen interface travelling towards the pores. The higher cooling rate assumes that none of the liquid nitrogen is vaporized or that any vaporized liquid nitrogen would travel instantly away from the surface by the action of buoyancy forces, while the lower cooling rate assumes that once the water touches the liquid nitrogen, a super cooled layer of gaseous nitrogen is formed. Using the mathematical model developed for Newtonian cooling by natural convection, and assuming that our heat transfer model resembles that of a vertical flat surface getting cooled by a fluid [89,90], we arrive at two possible cooling rates.

First, the heat transfer coefficient is calculated:

$$Nu = 0.27(Gr_L Pr)^{\frac{1}{4}} \qquad (1)$$

Where

$$Pr = \frac{\nu}{\alpha} = \frac{C_p \cdot \mu}{k} \qquad (2)$$

$$Gr_L = \frac{g\beta}{\nu^2}(T_o - T_\infty)L^3 \qquad (3)$$

$$h = Nu_L \frac{k}{L} \qquad (4)$$

We will assume that the cooling of the water film is Newtonian, as the following mathematical condition is satisfied:

$$Bi = \frac{hL}{k} \leq 0.1 \tag{5}$$

Thus, the temperature of the water at different times could be analyzed through the following relation,

$$\frac{T-T_f}{T_i-T_f} = exp\left[\frac{-hAt}{\rho C_p V}\right] \tag{6}$$

Though it is justified to assume Newtonian cooling, the solution above does eliminate the contribution of convective currents within the thin water layer. Also, the conduction from the metallic holder is neglected. So future simulations will be dedicated to taking all such factors into consideration for a better understanding of the cooling process throughout the water component of the system. Variable definitions, values used for this calculation and approximated cooling rates are in Table S1. Schematic for freezing experiment is also included in Figure S11.

**Specimen handling/transfer.** After final milling, the cryo-prepared specimens were transferred from the PFIB chamber to a side chamber which maintained in an ultra-high vacuum and at approx. -160 $_o$C, for less than 15 sec, and then into the pre-cooled ultra-high vacuum carry suitcase, maintained at approx. -190 $_o$C. The suitcase is then detached from the PFIB and mounted onto the Cameca LEAP 5000XS system. The cold puck is then transferred under cryo-UHV conditions to the atom probe analysis chamber.

**APT.** APT analysis was conducted using a Cameca local electrode atom probe (LEAP) 5000 XS (Cameca Instruments, USA). Data were acquired while operating in laser-pulsing mode with a pulse of 20-100 pJ and a pulse rate of 25-200 kHz. The target evaporation rate was set to 0.003 or 0.005 ions / pulse (0.3 or 0.5 %) by adjusting an applied DC voltage (typically ranging from approximately 2-5 kV). An example of ice APT specimen alignment to atom probe local (counter) electrode and the measurement are shown in Figure S8. The base temperature for the specimen stage was set 70 K throughout the measurement and the chamber pressure was in the 10$_{-11}$ torr range. Reconstruction and analysis were performed using Imago Visualization and Analysis Software (IVAS) 3.8.4, using SEM images of the tips to assist with spatial calibration of the reconstructions, as well as custom routines in MATLAB software.

# References

1. Nobel Prize ®and the Nobel Prize ®medal design mark are registered trademarks of the Nobel Foundation Scientific Background on the Nobel Prize in Chemistry 2017 THE DEVELOPMENT OF CRYO-ELECTRON MICROSCOPY. (2017).
2. Liao, H.-G. & Zheng, H. Liquid Cell Transmission Electron Microscopy. *Annu. Rev. Phys. Chem.* **67**, 719–747 (2016).
3. Park, J. *et al.* Nanoparticle imaging. 3D structure of individual nanocrystals in solution by electron microscopy. *Science* **349**, 290–295 (2015).
4. Chen, C.-C. *et al.* Three-dimensional imaging of dislocations in a nanoparticle at atomic resolution. *Nature* **496**, 74–77 (2013).
5. Yang, Y. *et al.* Deciphering chemical order/disorder and material properties at the single-atom level. *Nature* **542**, 75–79 (2017).
6. Zhou, J. *et al.* Observing crystal nucleation in four dimensions using atomic electron tomography. *Nature* **570**, 500–503 (2019).
7. Lefebvre-Ulrikson, W., Vurpillot, F. & Sauvage, X. *Atom probe tomography : put theory into practice*. (Academic Press, 2016).
8. De Geuser, F. & Gault, B. Metrology of small particles and solute clusters by atom probe tomography. *Acta Mater.* **188**, 406–415 (2020).
9. Perea, D. E., Gerstl, S. S. A., Chin, J., Hirschi, B. & Evans, J. E. An environmental transfer hub for multimodal atom probe tomography. *Adv. Struct. Chem. imaging* **3**, 12 (2017).
10. Stephenson, L. T. *et al.* The Laplace project: an integrated suite for correlative atom probe tomography and electron microscopy under cryogenic and UHV conditions. *PLoS One* **13**, e0209211 (2018).
11. Schreiber, D. K., Perea, D. E., Ryan, J. V., Evans, J. E. & Vienna, J. D. A method for site-specific and cryogenic specimen fabrication of liquid/solid interfaces for atom probe tomography. *Ultramicroscopy* (2018) doi:10.1016/j.ultramic.2018.07.010.
12. Stintz, A. & Panitz, J. A. Imaging Atom-Probe Analysis of an Aqueous Interface. *J. Vac. Sci. Technol. a-Vacuum Surfaces Film.* **9**, 1365–1367 (1991).
13. Stintz, A. & Panitz, J. A. Isothermal Ramped Field-Desorption of Water from Metal-Surfaces. *J. Appl. Phys.* **72**, 741–745 (1992).
14. Pinkerton, T. D. *et al.* Electric Field Effects in Ionization of Water-Ice Layers on Platinum.

*Langmuir* **15**, 851–855 (1999).

15. Schwarz, T., Stender, P., Schmitz, G. Fundamental Study of Cryo-FIB Produced Pure Water Tips by Atom Probe Tomography. in *1st virtual NRW-APT user meeting* (2020).
16. Qiu, S. *et al.* Three-Dimensional Chemical Mapping of a Single Protein in the Hydrated State with Atom Probe Tomography. *Anal. Chem.* **92**, 5168–5177 (2020).
17. Panitz, J. A. In search of the chimera: Molecular imaging in the atom-probe. in *Microscopy and Microanalysis* vol. 11 92–93 (2005).
18. Panitz, J. A. Point-projection imaging of unstained ferritin clusters. *Ultramicroscopy* **7**, 241–248 (1982).
19. Narayan, K., Prosa, T. J., Fu, J., Kelly, T. F. & Subramaniam, S. Chemical mapping of mammalian cells by atom probe tomography. *J. Struct. Biol.* **178**, 98–107 (2012).
20. Prosa, T., Kostrna Keeney, S. & Kelly, T. F. Local electrode atom probe analysis of poly(3-alkylthiophene)s. *J. Microsc.* **237**, 155–167 (2010).
21. Perea, D. E. *et al.* Atom probe tomographic mapping directly reveals the atomic distribution of phosphorus in resin embedded ferritin. *Sci. Rep.* **6**, 1–9 (2016).
22. Rusitzka, K. A. K. *et al.* A near atomic-scale view at the composition of amyloid-beta fibrils by atom probe tomography. *Sci. Rep.* (2018) doi:10.1038/s41598-018-36110-y.
23. Gault, B. *et al.* Investigation of Self-assembled Monolayer by Atom Probe Microscopy. *Microsc. Microanal.* **15**, 272–273 (2009).
24. Nickerson, B. S., Karahka, M. & Kreuzer, H. J. Disintegration and field evaporation of thiolate polymers in high electric fields. *Ultramicroscopy* **159**, 173–177 (2015).
25. Eder, K. *et al.* A New Approach to Understand the Adsorption of Thiophene on Different Surfaces: An Atom Probe Investigation of Self-Assembled Monolayers. *Langmuir* **33**, 9573–9581 (2017).
26. Li, T. *et al.* Atomic Imaging of Carbon-Supported Pt, Pt/Co, and Ir@Pt Nanocatalysts by Atom-Probe Tomography. *ACS Catal.* **4**, 695–702 (2014).
27. Tedsree, K. *et al.* Hydrogen production from formic acid decomposition at room temperature using a Ag–Pd core–shell nanocatalyst. *Nat. Nanotechnol.* **6**, 302–307 (2011).
28. Felfer, P., Benndorf, P., Masters, A., Maschmeyer, T. & Cairney, J. M. Revealing the distribution of the atoms within individual bimetallic catalyst nanoparticles. *Angew. Chem. Int. Ed. Engl.* **53**, 11190–11193 (2014).

29. Kim, S. H. *et al.* Characterization of Pd and Pd@Au core-shell nanoparticles using atom probe tomography and field evaporation simulation. *J. Alloys Compd.* **831**, 154721 (2020).

30. Perea, D. E. *et al.* Three-dimensional nanoscale composition mapping of semiconductor nanowires. *Nano Lett.* **6**, 181–185 (2006).

31. Du, S. *et al.* Quantitative dopant distributions in GaAs nanowires using atom probe tomography. *Ultramicroscopy* **132**, 186–192 (2013).

32. Lim, J. *et al.* Atomic-Scale Mapping of Impurities in Partially Reduced Hollow $TiO_2$ Nanowires. *Angew. Chemie - Int. Ed.* (2020) doi:10.1002/anie.201915709.

33. Diercks, D. R. *et al.* Atom probe tomography evaporation behavior of C-axis GaN nanowires: Crystallographic, stoichiometric, and detection efficiency aspects. *J. Appl. Phys.* **114**, 184903 (2013).

34. Kim, S.-H. *et al.* Direct Imaging of Dopant and Impurity Distributions in 2D $MoS_2$. *Adv. Mater.* (2020) doi:10.1002/adma.201907235.

35. Felfer, P. *et al.* New approaches to nanoparticle sample fabrication for atom probe tomography. *Ultramicroscopy* **159**, 413–419 (2015).

36. Kim, S.-H. *et al.* A new method for mapping the three-dimensional atomic distribution within nanoparticles by atom probe tomography (APT). *Ultramicroscopy* **190**, 30–38 (2018).

37. Kim, S. H., Lee, J. Y., Ahn, J. P. & Choi, P. P. Fabrication of Atom Probe Tomography Specimens from Nanoparticles Using a Fusible Bi-In-Sn Alloy as an Embedding Medium. *Microsc. Microanal.* (2019) doi:10.1017/S1431927618015556.

38. El-Zoka, A. A., Langelier, B., Botton, G. A. & Newman, R. C. Enhanced analysis of nanoporous gold by atom probe tomography. *Mater. Charact.* **128**, (2017).

39. Wayne, R. P. (Richard P. *Chemistry of atmospheres : an introduction to the chemistry of the atmospheres of earth, the planets, and their satellites*. (Oxford University Press, 2000).

40. Sprenger, M. *et al.* The Demographics of Water: A Review of Water Ages in the Critical Zone. *Reviews of Geophysics* (2019) doi:10.1029/2018RG000633.

41. Duley, W. W. Molecular Clusters in Interstellar Clouds. *Astrophys. J.* (1996) doi:10.1086/310326.

42. Khusnatdinov, N. N., Petrenko, V. F. & Levey, C. G. Electrical properties of the ice/solid interface. *J. Phys. Chem. B* **101**, 6212–6214 (1997).

43. Newman, R. C. 2.05 - Dealloying. in *Shreir's Corrosion* 801–809 (2010). doi:http://dx.doi.org/10.1016/B978-044452787-5.00031-7.

44. Xue, Y., Markmann, J., Duan, H., Weissmüller, J. & Huber, P. Switchable imbibition in nanoporous gold. *Nat. Commun.* **5**, (2014).

45. Raspal, V. *et al.* Nanoporous surface wetting behavior: The line tension influence. *Langmuir* **28**, 11064–11071 (2012).

46. Halpin, J. E. *et al.* An in-situ approach for preparing atom probe tomography specimens by xenon plasma-focussed ion beam. *Ultramicroscopy* (2019) doi:10.1016/j.ultramic.2019.04.005.

47. Loi, S. T., Gault, B., Ringer, S. P., Larson, D. J. & Geiser, B. P. Electrostatic simulations of a local electrode atom probe: The dependence of tomographic reconstruction parameters on specimen and microscope geometry. *Ultramicroscopy* (2013) doi:10.1016/j.ultramic.2012.12.012.

48. Miller, M. K., Russell, K. F. & Thompson, G. B. Strategies for fabricating atom probe specimens with a dual beam FIB. *Ultramicroscopy* **102**, 287–298 (2005).

49. Thompson, K. *et al.* In situ site-specific specimen preparation for atom probe tomography. *Ultramicroscopy* (2007) doi:10.1016/j.ultramic.2006.06.008.

50. BASSIM, N. D. *et al.* Minimizing damage during FIB sample preparation of soft materials. *J. Microsc.* **245**, 288–301 (2012).

51. Marko, M., Hsieh, C., Schalek, R., Frank, J. & Mannella, C. Focused-ion-beam thinning of frozen-hydrated biological specimens for cryo-electron microscopy. *Nat. Methods* **4**, 215–217 (2007).

52. Chang, Y. H. *et al.* Quantification of solute deuterium in titanium deuteride by atom probe tomography with both laser pulsing and high-voltage pulsing: Influence of the surface electric field. *New J. Phys.* (2019) doi:10.1088/1367-2630/ab1c3b.

53. Panitz, J. A. & Stintz, A. Imaging atom-probe analysis of a vitreous ice interface. *Surf. Sci.* (1991) doi:10.1016/0039-6028(91)90408-K.

54. Carrasco, J., Hodgson, A. & Michaelides, A. A molecular perspective of water at metal interfaces. *Nat. Mater.* **11**, 667–674 (2012).

55. Silaeva, E. P. *et al.* Do dielectric nanostructures turn metallic in high-electric dc fields? *Nano Lett.* **14**, 6066–6072 (2014).

56. Kelly, T. F. *et al.* Laser pulsing of field evaporation in atom probe tomography. *Curr. Opin. Solid State Mater. Sci.* **18**, 81–89 (2014).

57. Block, J.H. & Cocke, D. L. Field Ion and Field Desorption Mass Spectrometry of Inorganic Compounds. *Surf. Sci.* **70**, 363–391 (1978).

58. Karahka, M. & Kreuzer, H. J. Water whiskers in high electric fields. in *Physical Chemistry Chemical Physics* vol. 13 11027–11033 (The Royal Society of Chemistry, 2011).

59. Bluhm, J., Ricken, T. & Bloßfeld, M. Ice Formation in Porous Media BT - Advances in Extended and Multifield Theories for Continua. in (ed. Markert, B.) 153–174 (Springer Berlin Heidelberg, 2011).

60. Bazant, M. Z. Theory of ElectroChemical Kinetics based on Nonequilibrium Thermodynamics. *ArXiv e-prints* (2013).

61. Ginot, F., Lenavetier, T., Dedovets, D., Deville, S. Solute strongly impacts freezing under confinement. *Phys, Appl* **116**, (2020).

62. Fumagalli, L. *et al.* Anomalously low dielectric constant of confined water. *Science (80-. ).* **360**, 1339–1342 (2018).

63. Viskanta, R., Bianchi, M. V. A., Critser, J. K. & Gao, D. Solidification processes of solutions. *Cryobiology* (1997) doi:10.1006/cryo.1997.2015.

64. Körber, C., Scheiwe, M. W. & Wollhöver, K. Solute polarization during planar freezing of aqueous salt solutions. *Int. J. Heat Mass Transf.* (1983) doi:10.1016/s0017-9310(83)80179-3.

65. El-Zoka, A. A., Langelier, B., Korinek, A., Botton, G. A. & Newman, R. C. Nanoscale mechanism of the stabilization of nanoporous gold by alloyed platinum. *Nanoscale* **10**, (2018).

66. Tsong, T. T. Formation of silicon and water cluster ions in pulsed-laser stimulated field desorption. *J. Vac. Sci. Technol. B Microelectron. Nanom. Struct.* **3**, 1425 (1985).

67. Larson, D. J. J., Gault, B., Geiser, B. P. P., De Geuser, F. & Vurpillot, F. Atom probe tomography spatial reconstruction: Status and directions. *Curr. Opin. Solid State Mater. Sci.* **17**, 236–247 (2013).

68. Wang, X. *et al.* Interpreting nanovoids in atom probe tomography data for accurate local compositional measurements. *Nat. Commun.* **11**, (2020).

69. Stuve, E. M. Ionization of water in interfacial electric fields: An electrochemical view.

*Chem. Phys. Lett.* **519–520**, 1–17 (2012).

70. Stintz, A. & Panitz, J. A. Cluster ion formation in isothermal ramped field-desorption of amorphous water ice from metal surfaces. *Surf. Sci.* **296**, 75–86 (1993).
71. Pennsylvania, T. Cluster-ion formation. **32**, 4340–4357 (1985).
72. Müller, M., Saxey, D. W., Smith, G. D. W. & Gault, B. Some aspects of the field evaporation behaviour of GaSb. *Ultramicroscopy* **111**, 487–492 (2011).
73. Zanuttini, D. *et al.* Simulation of field-induced molecular dissociation in atom-probe tomography: Identification of a neutral emission channel. *Phys. Rev. A* **95**, 61401 (2017).
74. Peng, Z. *et al.* Unraveling the Metastability of $Cn2+$ (n = 2-4) Clusters. *J. Phys. Chem. Lett.* **10**, 581–588 (2019).
75. Blum, I. *et al.* Dissociation dynamics of molecular ions in high DC electric field. *J. Phys. Chem. A* **120**, 3654–3662 (2016).
76. Tsong, T. Pulsed-laser-stimulated field ion emission from metal and semiconductor surfaces: A time-of-flight study of the formation of atomic, molecular, and cluster ions. *Phys. Rev. B* **30**, 4946–4961 (1984).
77. Qiu, S. *et al.* Graphene encapsulation enabled high-throughput atom probe tomography of liquid specimens. *Ultramicroscopy* 113036 (2020) doi:10.1016/j.ultramic.2020.113036.
78. Rusitzka, K. A. K. *et al.* A near atomic-scale view at the composition of amyloid-beta fibrils by atom probe tomography. *Sci. Rep.* **8**, (2018).
79. Wang, L. R. C., Kreuzer, H. J. & Nishikawa, O. Polythiophene in strong electrostatic fields. *Org. Electron.* **7**, 99–106 (2006).
80. Nishikawa, O. & Kato, H. Atom-probe study of a conducting polymer: The oxidation of polypyrrole. *J. Chem. Phys.* **85**, 6758–6764 (1986).
81. Forbes, R. G. Field evaporation theory: a review of basic ideas. *Appl. Surf. Sci.* **87–88**, 1–11 (1995).
82. Driesel, W., Dietzsch, C., Niedrig, H. & Praprotnik, B. HV TEM in situ investigations of the tip shape of a gallium liquid-metal ion/electron emitter. *Ultramicroscopy* **57**, 45–58 (1995).
83. Gault, B., Danoix, F., Hoummada, K., Mangelinck, D. & Leitner, H. Impact of directional walk on atom probe microanalysis. *Ultramicroscopy* **113**, 182–191 (2012).
84. Katnagallu, S. *et al.* Impact of local electrostatic field rearrangement on field ionization. *J.*

*Phys. D. Appl. Phys.* **51**, 105601 (2018).

85. De Geuser, F., Gault, B., Bostel, A. & Vurpillot, F. Correlated field evaporation as seen by atom probe tomography. *Surf. Sci.* **601**, 536–543 (2007).
86. Gault, B. *et al.* Behavior of molecules and molecular ions near a field emitter. *New J. Phys.* **18**, 33031 (2016).
87. Feistel, R. & Wagner, W. Sublimation pressure and sublimation enthalpy of H2O ice Ih between 0 and 273.16 K. *Geochim. Cosmochim. Acta* (2007) doi:10.1016/j.gca.2006.08.034.
88. El-Zoka, A. A., Howe, J. Y., Newman, R. C. & Perovic, D. D. In situ STEM/SEM Study of the Coarsening of Nanoporous Gold. *Acta Mater.* (2018) doi:10.1016/j.actamat.2018.09.002.
89. Poirier, D. R. & Geiger, G. H. *Transport phenomena in materials processing*. *Transport Phenomena in Materials Processing* (2016). doi:10.1007/978-3-319-48090-9.
90. McAdams, W. H. *Heat Transmission*. (McGraw-Hill, 1954).
91. M.W. Chase Jr. *NIST-JANAF Themochemical Tables, Fourth Edition*. *J. Phys. Chem. Ref. Data* (1998).
92. Jensen, J. E., Tuttle, W. A., Stewart, R. B., Brechna, H. & Prodell, A. G. Properties of Nitrogen. *Brookhaven Natl. Lab. Sel. Cryog. Data Noteb.* (1980).

# Supplementary Information

# Enabling near atomic-scale analysis of liquids

A.A. El-Zoka, S.-H. Kim, L.T. Stephenson, B. Gault

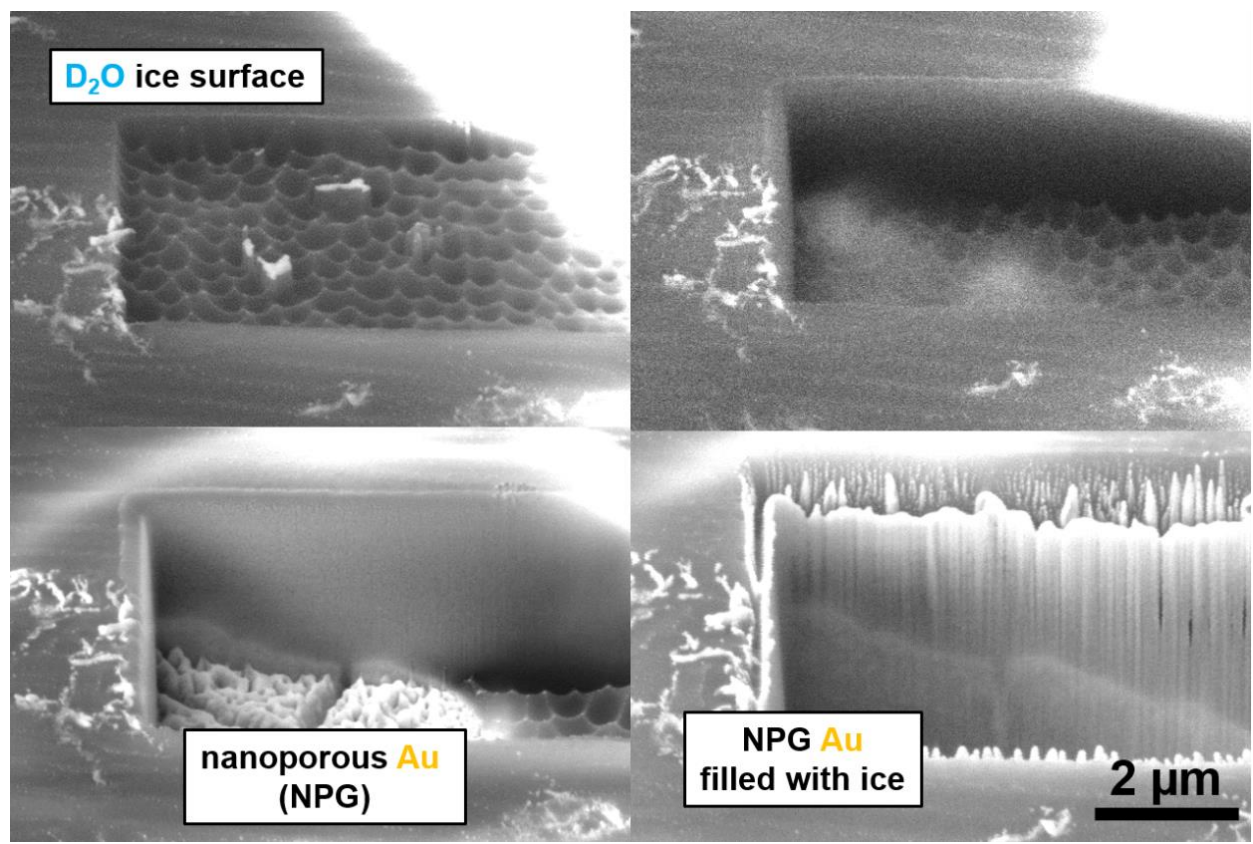

**Figure S1.** Xe-plasma focused ion beam (PFIB)- secondary electron microscopy (SEM) images of cross-sections of a $D_2O$ ice layer and nanoporous gold (NPG). The result clearly shows that the NPG structure enhanced wettability by $D_2O$.

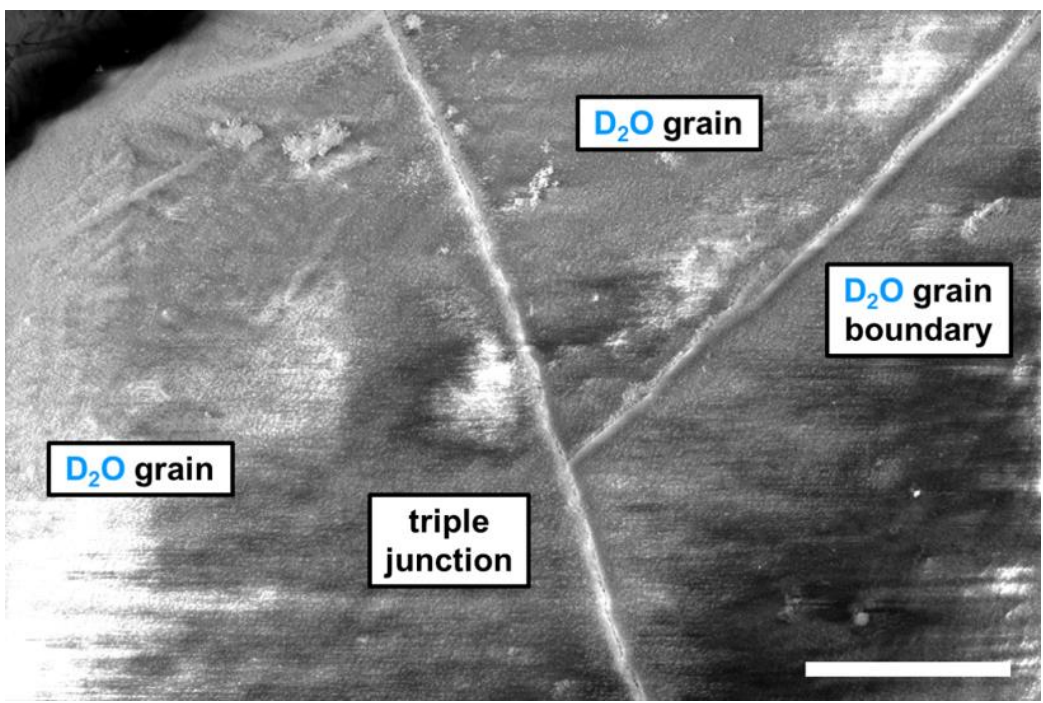

**Figure S2.** SEM images of the surface of the sample at regions of higher thicknesses of ice on top of the NPG, showing crystallinity features such as grain boundaries and triple junctions. Scale bar is 50 μm.

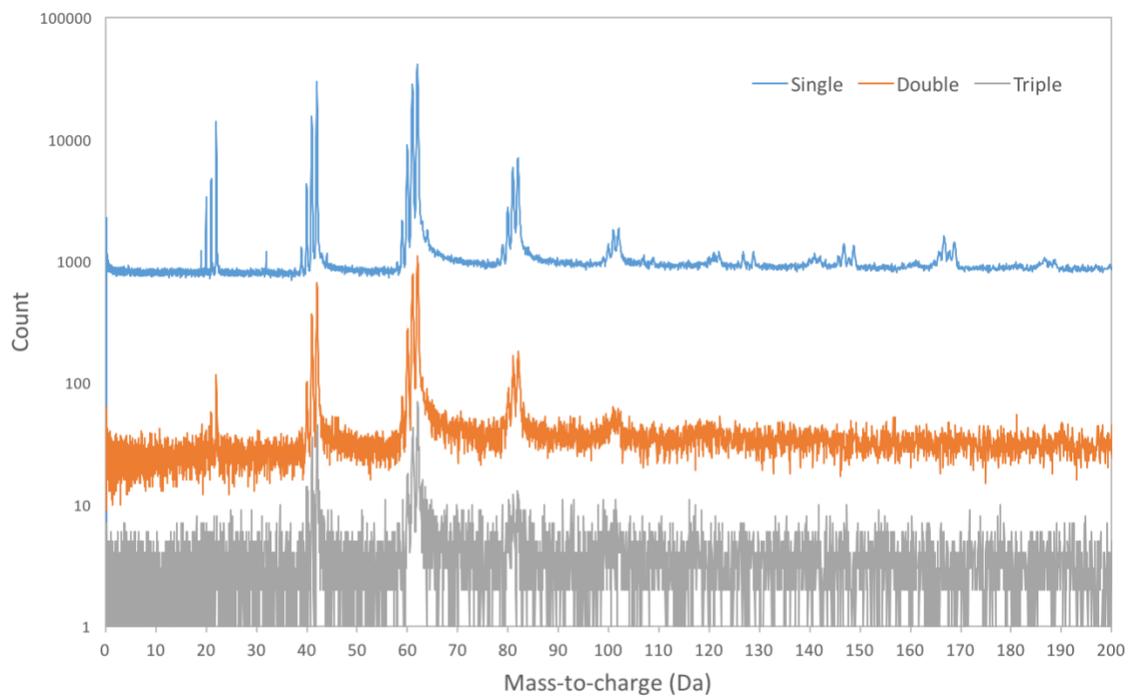

**Figure S3.** Mass spectra obtained from after filtering for multiples hits: singles (blue), doubles (orange), and triples (grey). The sample was run at laser energy of 40 pJ and pulse frequency of 200 kHz.

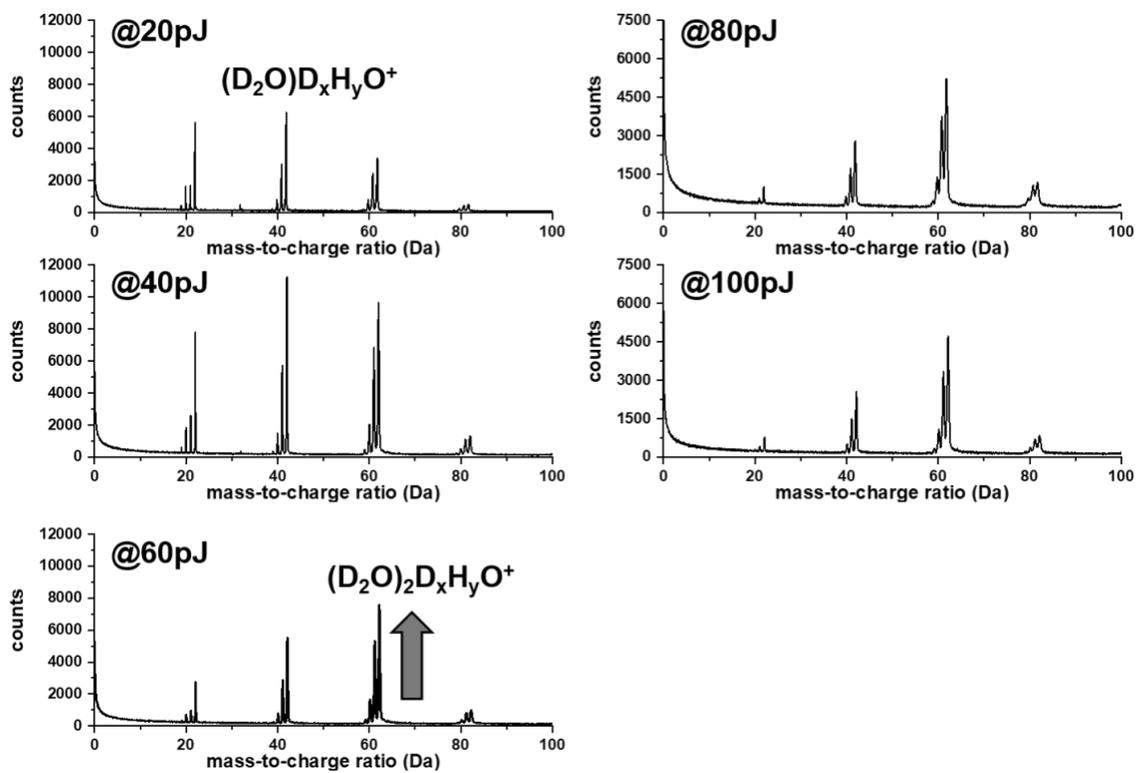

**Figure S4.** Mass spectra of ice from APT measurement at different pulsed laser energies.

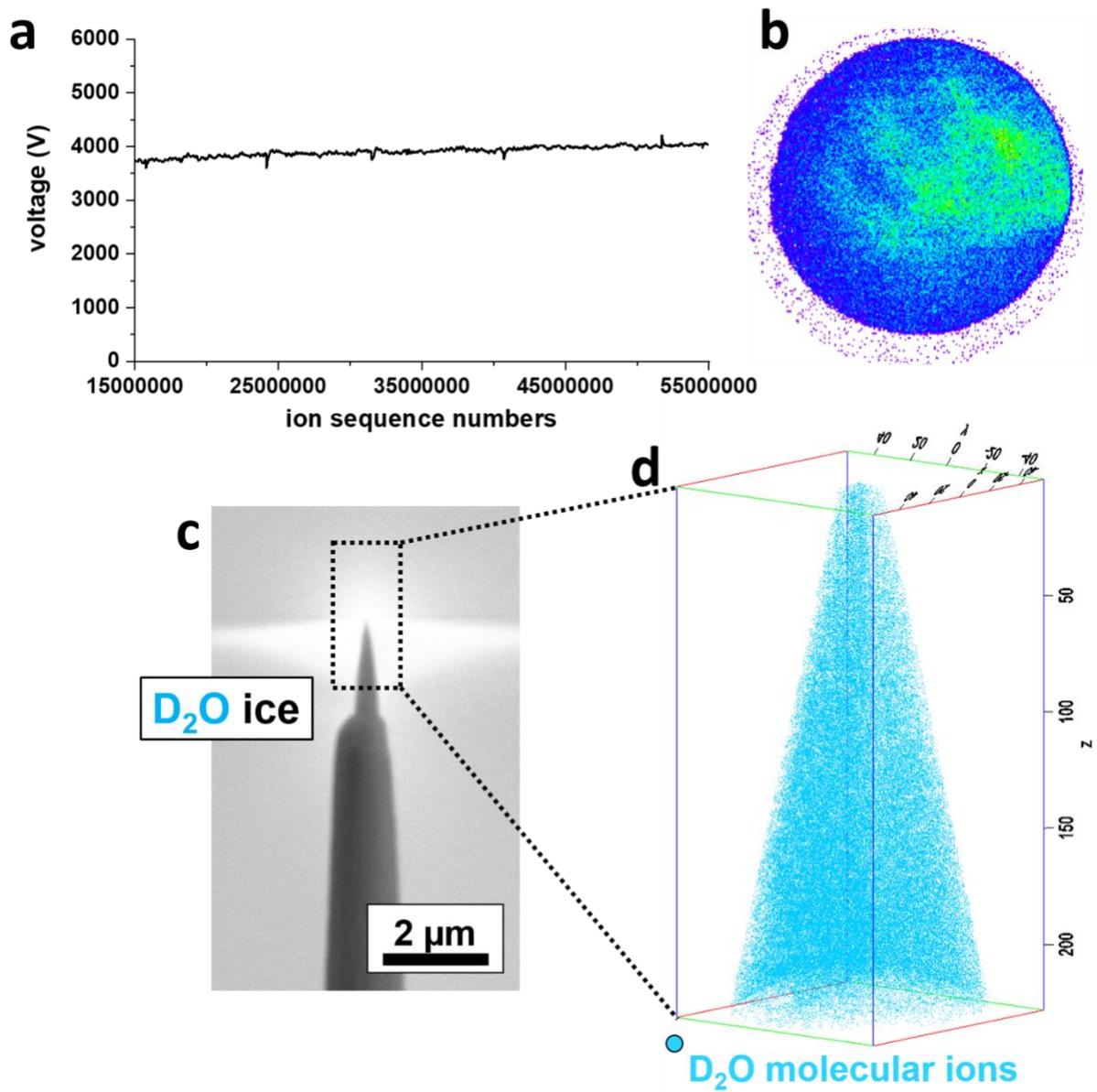

**Figure S5.** (a) Voltage history of the APT measurement and (b) corresponding detector histogram. (c) SEM image of APT specimen of ice and (d) corresponding 3D reconstructed atom map of $D_2O$ (scale is in nm).

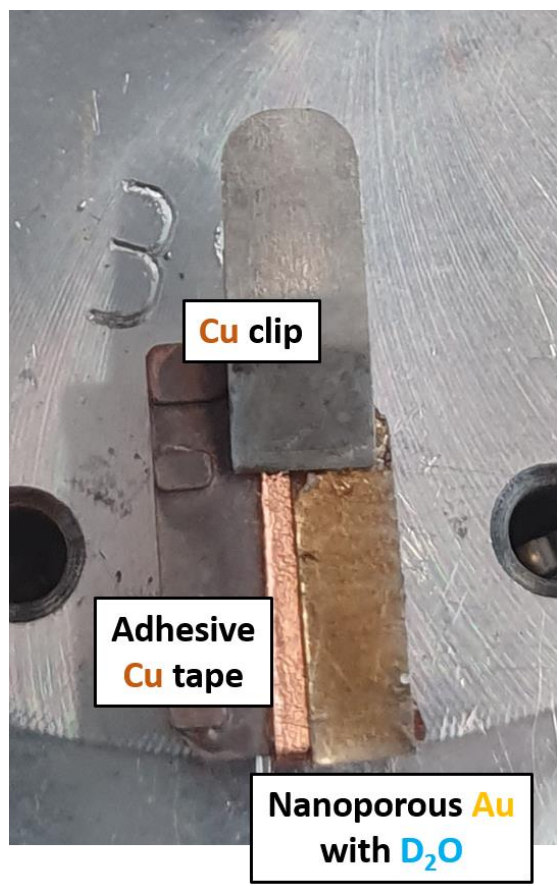

**Figure S6.** NPG foil mounted on a commercial LEAP system Cu clip with adhesive Cu tape. The NPG foil was immersed into D$_2$O overnight.

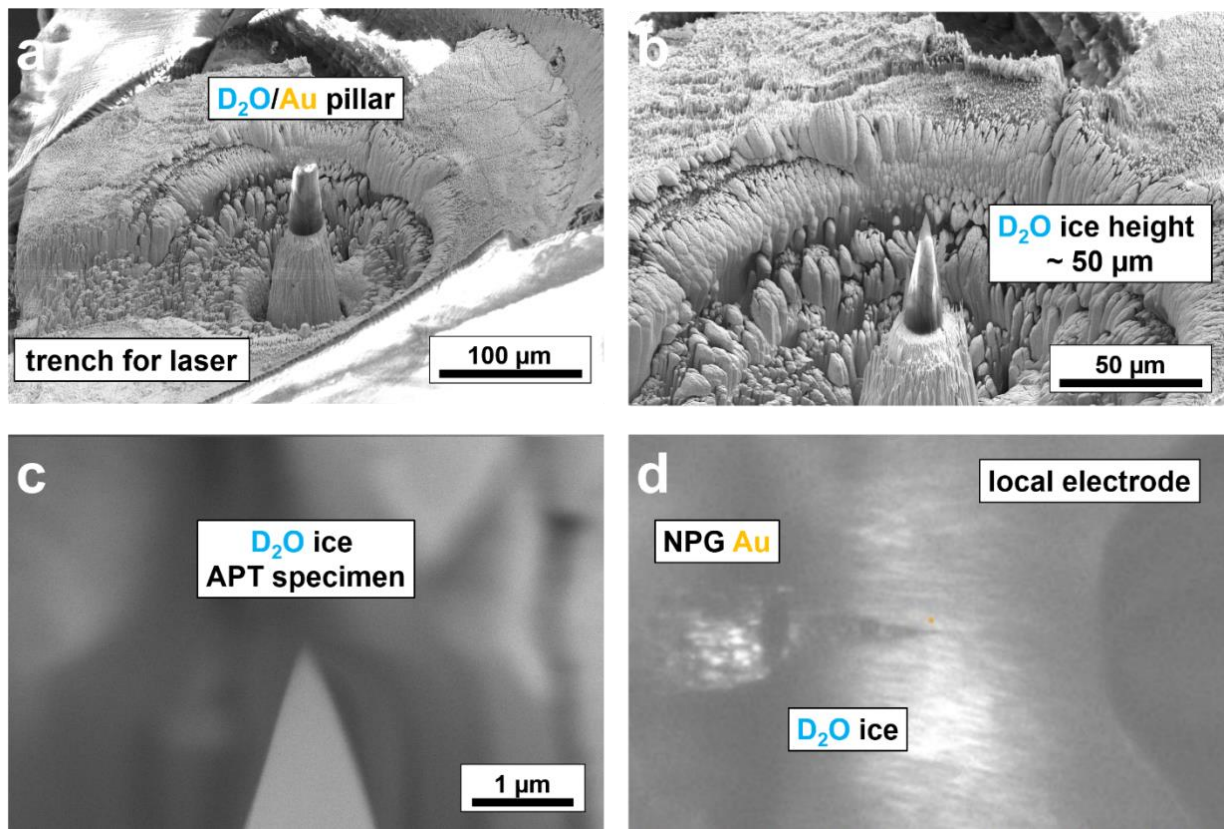

**Figure S7.** (a-c) SEM images of APT specimen preparation with sample size of >50 µm. (d) optical microscopy image of the sample aligned with the atom probe local electrode.

Frozen $D_2O$ on a NPG substrate was sharpened into an APT specimen with >50 µm in height. Sample fabrication and transfer to the atom probe instrument were successful; however, no ions were detected in either pulsed voltage or laser mode. As ice is an insulator and the specimen length is too long, the charge density on the apex is too low to allow evaporation, and eventually the sample fractured at the interface between $D_2O$ and NPG.

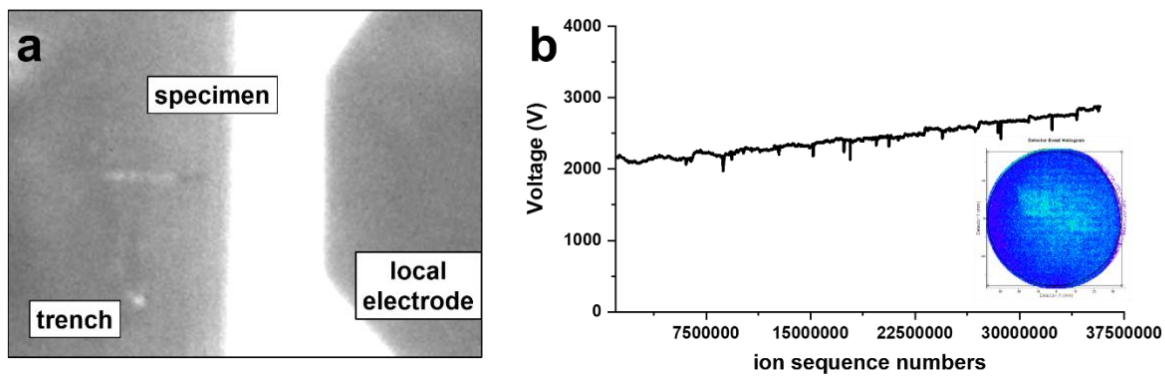

**Figure S8.** (a) Optical microscopy image of an ice APT specimen showing the local electrode and trenches from Xe-ion milling. (b) Voltage history of ice APT measurement. The inset image is the corresponding ion detector histogram. Note that the reconstruction of this sample is shown in Figure 2(c).

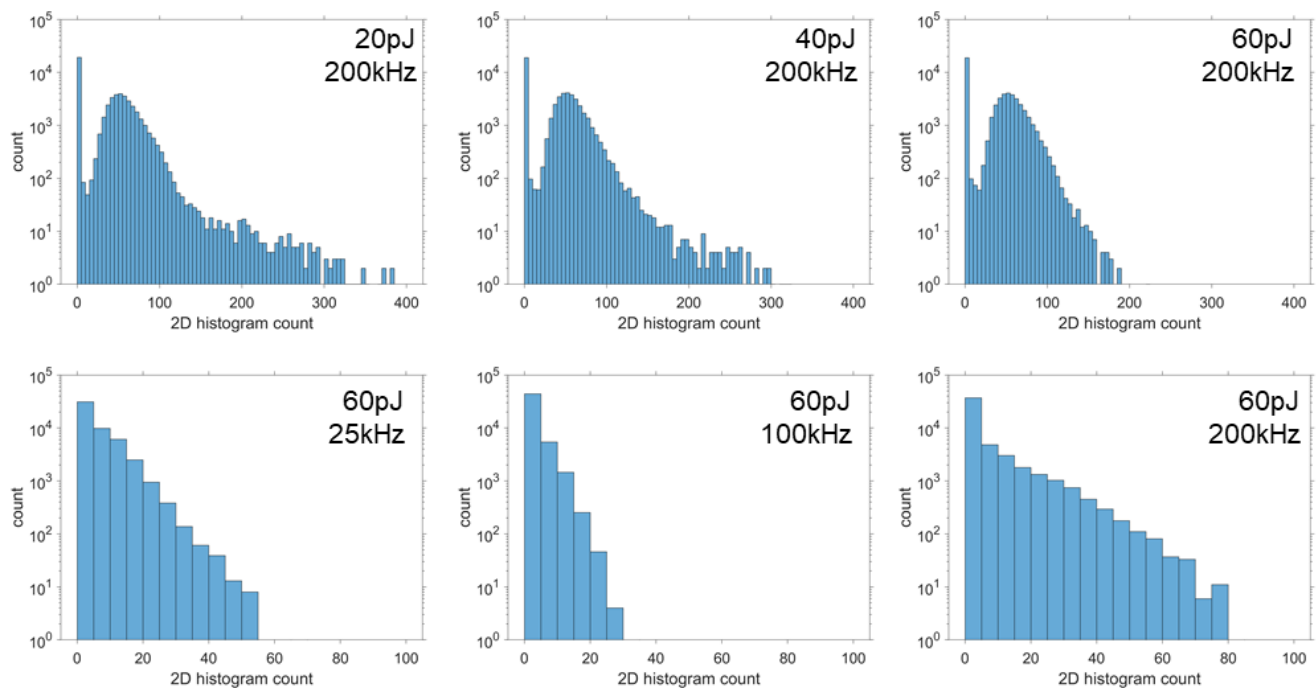

**Figure S9.** 2D distributions of intensities detected across the detector maps for a set of pulsing energy/pulsing rate parameters used in Figure 8.

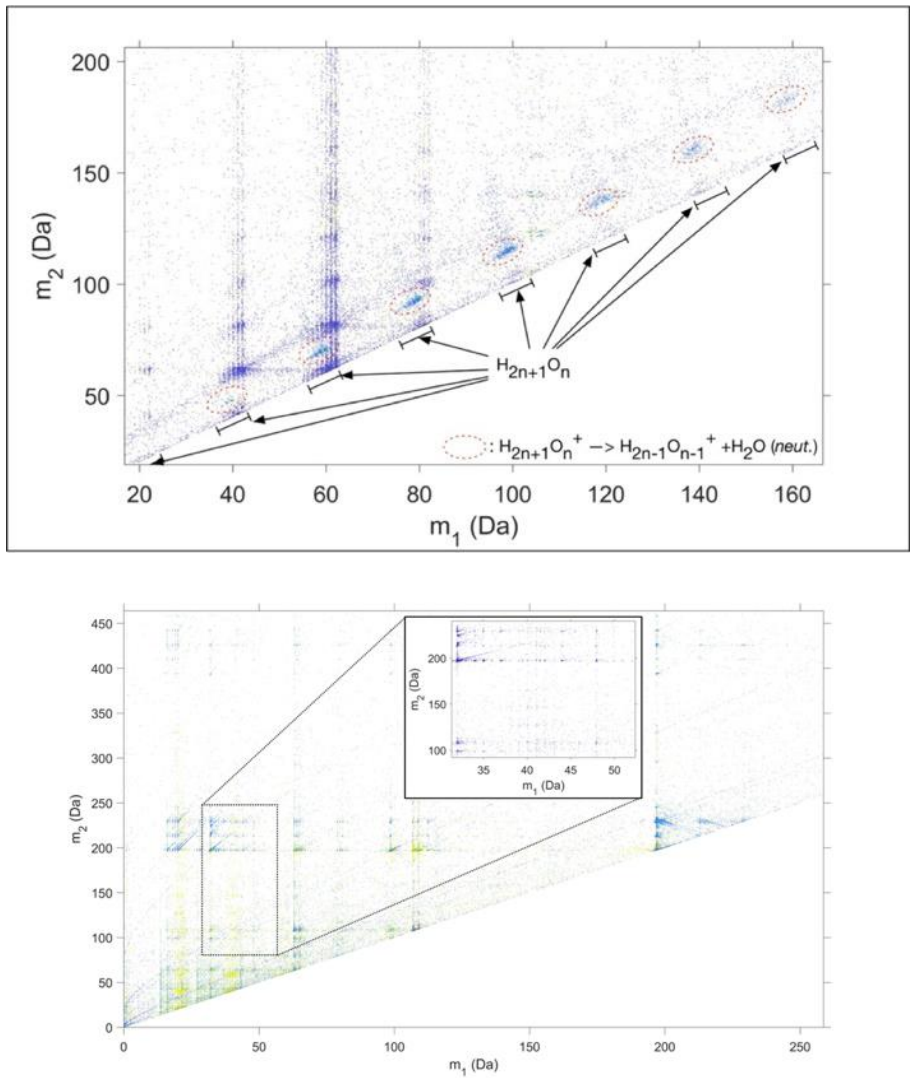

**Figure S10**. Mass-to-charge correlation plots for multiple events. (a) Multiple evaporation correlation plot for a 10mM NaCl water, displaying the late dissociation of water clusters involving one detected neutral water molecule being released. (b) Multiple evaporation correlation plot for a 10mM NaCl water upon ice-NPG composite, displaying multiple co-evaporation of various species including dissociation of $Au_2(OH)_2^{2+}$, $Au_2O_2^{2+}$ and other complex species. Insert displays some co-evaporation which could be explained by $Cl^+$ (35 & 37 Da) and $ClO_3^{2+}$ (41.5 & 43.5 Da) ions but these signals are tiny and more in-depth classification is required.

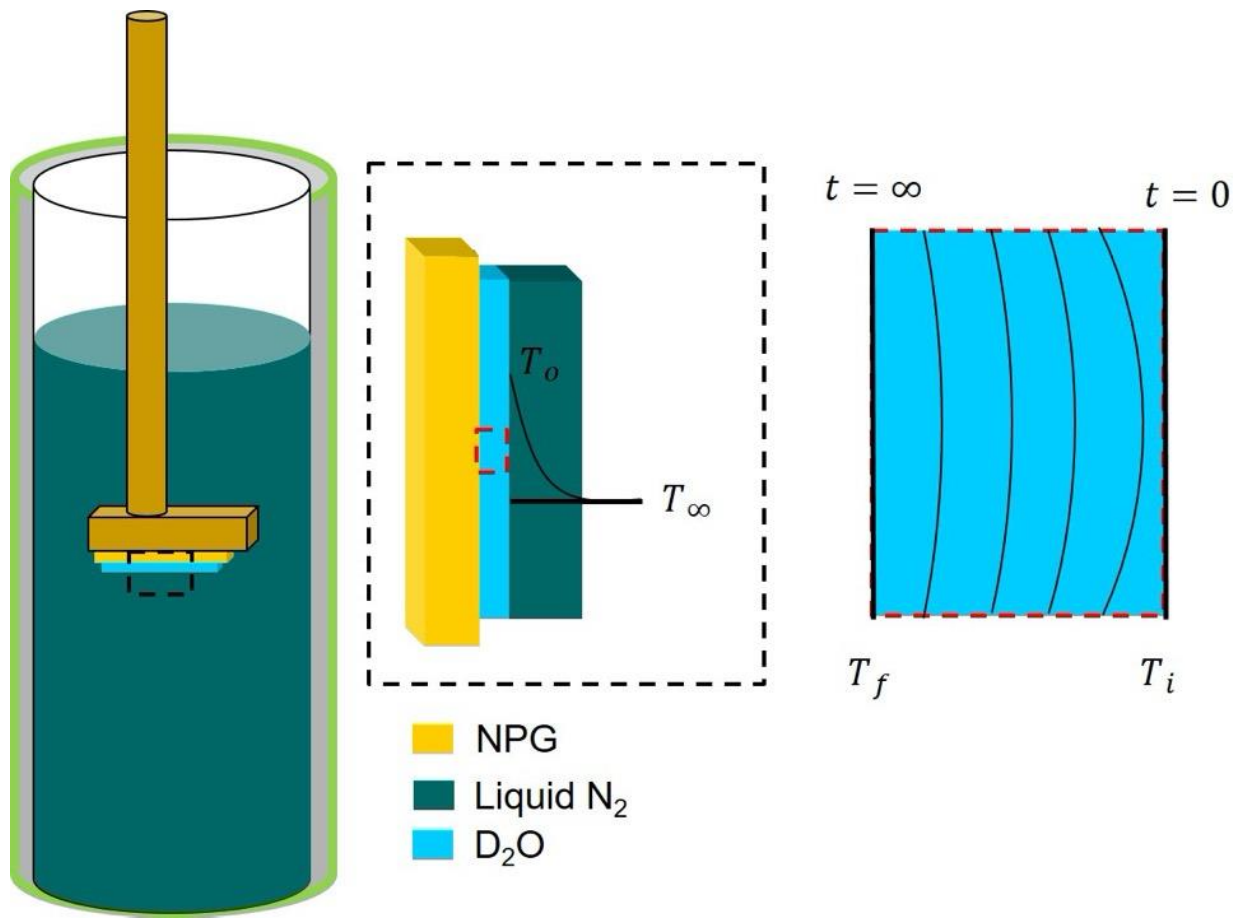

**Figure S11.** Schematic showing the setup for quenching NPG/water samples in liquid nitrogen. Insets show the temperature profiles at the water/liquid nitrogen interface and within the water layer on top of the NPG, in conformation with Newtonian cooling by natural convection.

**Table S1- List of variable definitions and values** [91,92] **used in approximating cooling rates.**

| Symbol | Variable/Number | Units | Value Implemented (Lowest Cooling Rate) | Value Implemented (Highest Cooling Rate) |
|---|---|---|---|---|
| Nu | Nusselt number | -- | 4.99E+01 | 2.05E+02 |
| h | Heat transfer rate | W.m$^{-2}$K$^{-1}$ | 3.86E+01 | 2.66E+03 |
| L | Overhead length of water layer | m | 1.00E-02 | 1.00E-02 |
| k | Thermal conductivity of nitrogen | W.m$^{-1}$K$^{-1}$ | 7.74E-03 | 1.30E-01 |
| Pr | Prandtl number | -- | 7.32E-01 | 2.00E+00 |
| Bi | Biot number | -- | 6.46E-01 | 4.45E+01 |
| g | Acceleration due to earth's gravity | m$^2$s$^{-1}$ | 9.81E+00 | 9.81E+00 |
| β | Thermal expansion coefficient | K$^{-1}$ | 1.00E+00 | 2.00E+00 |
| T$_o$ | Surface Temperature | K | 2.98E+02 | 2.98E+02 |
| T$_\infty$ | Bulk Fluid Temperature | K | 8.30E+01 | 8.30E+01 |
| A | Area | m$^2$ | 1.00E-04 | 1.00E-04 |
| V | Volume | m$^3$ | 5.00E-10 | 5.00E-10 |
| Cp | Specific heat capacity | J.K$^{-1}$kg$^{-1}$ | 1.04E+03 | 2.00E+03 |
| T$_i$ | Initial Temperature | K | 2.98E+02 | 2.98E+02 |
| T | Target temperature | K | 1.36E+02 | 1.36E+02 |
| T$_f$ | Fluid Temperature | K | 8.30E+01 | 8.30E+01 |
| ρ | Density | kg.m$^{-3}$ | 9.97E+02 | 9.97E+02 |
| μ | Dynamic viscosity | Pa.s | 5.45E-06 | 1.30E-04 |
| Gr | Grashof number | -- | 1.59E+09 | 1.65E+11 |
| υ | Kinematic viscosity | m$^2$.s | 1.15E-06 | 1.60E-07 |
| k$_w$ | Thermal conductivity of water | W.m$^{-1}$K$^{-1}$ | 5.98E-01 | 5.98E-01 |
| t | Time | s | 1.88E-01 | 5.25E-03 |
| R | Cooling rate | K.s$^{-1}$ | 8.63E+02 | 3.09E+04 |